\documentclass[pdflatex,sn-nature]{sn-jnl}

\usepackage{graphicx}%
\usepackage{multirow}%
\usepackage{amsmath,amssymb,amsfonts}%
\usepackage{amsthm}%
\usepackage{mathrsfs}%
\usepackage[title]{appendix}%
\usepackage{xcolor}%
\usepackage{textcomp}%
\usepackage{manyfoot}%
\usepackage{booktabs}%
\usepackage{algorithm}%
\usepackage{algorithmicx}%
\usepackage{algpseudocode}%
\usepackage{listings}%
\usepackage{tabularx}
\usepackage{subcaption}
\usepackage{svg}
\usepackage[figuresright]{rotating}%

\begin{document}

\title[CES Network Paper]{Resilience of the reported global human-nature interaction network to pandemic conditions}

\author*[1]{\fnm{Anne Cathrine} \sur{Linder}}\email{acali@dtu.dk}

\author[1]{\fnm{David} \sur{Lusseau}}\email{davlu@dtu.dk}

\affil[1]{\orgdiv{National Institute of Aquatic Resources}, \orgname{Technical University of Denmark}, \orgaddress{\city{Kgs. Lyngby}, \postcode{2800}, \country{Denmark}}}

\abstract{Understanding human-nature interactions and the architecture of coupled human-nature systems is crucial for sustainable development. Cultural ecosystem services (CES), defined as intangible benefits derived from nature exposure, contribute to maintaining and improving human well-being. However, we have limited understanding of how well-being benefits emerge from CES co-production. In this study, for the first time, we estimated the global CES network from self-reported interactions between nature features and human activities underpinning CES co-production using social media. First, we used a bottom-up, replicable, approach to define the global repertoire of nature features and human activities used during CES co-production using 682,000 posts on Reddit. We then sampled Twitter to estimate the co-occurrence of these features and activities over the past five years, retrieving 41.7 millions tweets. These tweets were used to estimate the CES bipartite network, where each link was weighted by the number of times nature features and human activities co-occurred in tweets. We expected to observe large changes in the CES network topology in relation to the global mobility restrictions during the COVID-19 pandemic. This was not the case and the global CES network was generally resilient. However a higher order singular value decomposition of the number of tweets where a given feature and activity co-occur on a given day revealed an impulse on the link between self care activities and urban greenspace in response to the pandemic. This could be due to an increased need for self care during the pandemic and urban greenspace enabling CES to be produced locally. Thus, providing resilience for maintaining well-being during the pandemic. Our user based analysis also indicated a shift towards local CES production during the beginning of the pandemic, where there was a large increase in the proportion of users tweeting about specific CES features and activities for the first time. Thus, supporting that CES was produced locally. These findings, therefore, suggest an overall need for CES and access to features providing CES in local communities.}


\maketitle
\section{Introduction} \label{sec:intro}
One of the grand challenges of sustainability science is understanding human-nature interactions and the architecture and functions of coupled human-nature systems \cite{Kates2011, Clark2020}. This knowledge is crucial for sustaining human activities while keeping the planet habitable, at least. One poorly understood aspect of those interactions are cultural ecosystem services (CES). CES are generally defined as non-material benefits people obtain from nature \cite{Hirons2016}. CES, therefore, underpin one of the largest global business sectors (tourism) and contribute significantly to maintaining and improving human well-being \cite{TwohigBennett2018, Zhang2020}. When envisioning the production of CES, it is difficult to disentangle human activities from the natural settings in which they take place (and vice-versa) \cite{Fish2016, Raymond2018}. Consequently, in a socioecological paradigm, CES can be conceived as co-produced by human activities and nature features and emerge from people acting in nature \cite{Raymond2018}. However, we have limited understanding of how well-being benefits emerge from CES co-production. For example, it is not known which general ecosystem features are used for recreational activities and enjoyment nor which features and activities are important for eliciting well-being benefits \cite{Milcu2013, Hirons2016, Barton2016b, Mancini2019, Lai2019}.

Rapid planetary changes are placing significant burdens on both physical and mental wellbeing for a large section of the human species \cite{Palinkas2020}. CES can significantly contribute to countering this global well-being challenge.. Therefore, it is has become urgent to understand how co-production of CES may yield well-being benefits. In the socioecological paradigm of CES co-production, co-occurrence between human activities and nature features is dynamic. The resulting network of activities-nature features interactions has therefore the scope to adapt under changing conditions and change its topological state \cite{Kossinets2006, Gao2016}. Therefore, this complex adaptive system of interactions between nature features and human activities can exist in multiple states under different socioecological contexts and can exhibit different emergent properties associated with well-being maintenance and improvement.  

In 2020, the COVID-19 pandemic created an abrupt disruption to the daily life of individuals across the globe \cite{Mateer2021}. The pandemic led to reduced physical activity and lower mental health levels across multiple populations and countries \cite{Veldez2020}. Thus, there was a crucial need to counteract these negative psychological effects of the COVID-19 pandemic \cite{Hossain2020}. During the pandemic public health interventions meant that human mobility was globally reduced, hence reducing the scope for people to access nature. At the same time there are multiple reports that people sought nature more during the pandemic for its well-being benefits \cite{Venter2021, Lusseau2023}. Moreover, outdoor recreation and greenspace were suggested to foster well-being resilience by providing spaces to facilitate social interactions during the pandemic \cite{Mateer2021, Venter2021}. Thus, the global COVID-19 pandemic provides a natural experiment to assess the resilience and dynamics of the global human-nature interactions underpinning CES co-production.

User-generated data on social media can be a useful measure of visitation at destinations offering CES \cite{Mancini2019, daMota2020, Vigl2021}. We can also use text-mining approaches on those samples to retrieve the context of human-nature interactions from those online observations \cite{Erskine2021, Lusseau2023}. In this study, for the first time, we estimate the global CES network from self-reported interactions between ecosystem features and human activities underpinning CES co-production using social media. For the first time, we use a bottom-up, replicable, approach to define the global repertoire of ecosystem features and activities used during CES co-production using 682,000 posts on Reddit. We then estimated their co-occurrence over the past five years using 41.7 millions tweets. We hypothesized that CES networks, based on human-nature interactions, can be classified into distinguishable topological states and expected that the global mobility restrictions during the COVID-19 pandemic induced a state-shift in this network.

\section{Results} \label{sec:results}
\begin{figure}[H]
\centering
\includegraphics[ width=0.85\linewidth]{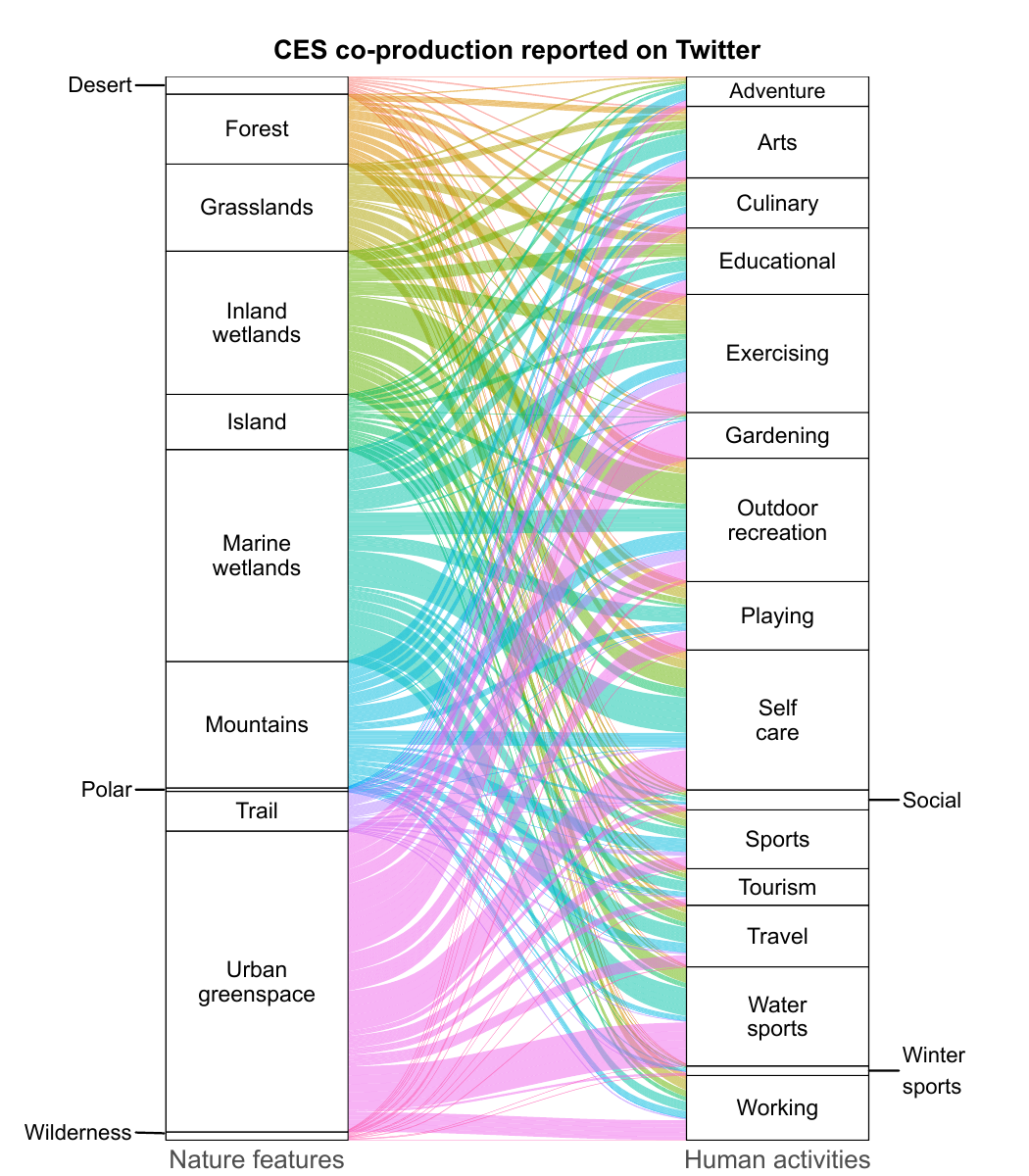}
\caption{Global CES network of tweets mentioning human-nature interactions on Twitter from 2018-2022 (41.7 million tweets). Links are weighted by the number of tweets including a given nature feature (grouped by nature class) and human activity (grouped by activity class).}
\label{fig:CES_tweet_network}
\end{figure}

We queried the subreddits r/Outdoors and r/EarthPorn over the past ten years to capture outdoor activities and nature features described in these subreddits that focus on CES. From the 682,206 posts retrieved we extracted the list of the 186 unique activities and 39 main nature features described in these posts (see \ref{sec:methods}). We then exhaustively sampled all English tweets mentioning the both one feature and one activity from 2018 to 2022 (41.7 million tweets after quality control procedures, see \ref{sec:methods}). We used these tweets to construct CES networks (Fig. \ref{fig:CES_tweet_network}). These are bipartite networks \cite{Newman2010}, where link are represented by the number of tweets where nature features (first type of node) co-occur with activities (second type of node). We estimated this network for all years combined (Fig. \ref{fig:CES_tweet_network}) and for each of the five years (Fig. \ref{fig:CES_annual_tweet_network}). Due to the large number of features and activities, a second network was constructed, where features and activities were pooled into larger classes, i.e., features were grouped into 11 nature classes and activities were grouped into 16 activity classes (see \ref{sec:methods}).
We also constructed a tensor representation of this network (\textbf{$CES_{f,a,d}$}), which regroup daily slices of the CES network: 
\[ CES_f,a,d=tweet_{f,a,d} \] i.e., the number of tweets retrieved where a given feature \textit{f} and activity \textit{a} co-occur in the text on a given day \textit{d} between 1 Jan 2018 and 31 Dec 2022.

There were no noticeable topological changes among the annual CES networks (Fig. \ref{fig:CES_annual_tweet_network}). The global network statistics that capture broad topological features, i.e., network asymmetry, modularity, nestedness, connectance, and interaction asymmetry, did not change between the years (Tab. \ref{tab:net_stats}). This was the case for both full and grouped representations of the CES networks (Tab. \ref{tab:net_stats}). The node level statistics, interaction asymmetry and nested rank also did not noticeably change across years (Tab. \ref{tab:node_stats}, Fig. \ref{fig:node_level_stats}). The interaction asymmetry did however reveal that the majority of activities with the exception of self care activities, outdoor recreation activities, and exercising activities were being pulled by the nature features, whereas, the majority of nature features were pushing the activities. Moreover, the nested rank identified self care activities and outdoor recreation activities to have the highest generality among the activities and urban greenspace features had the highest generality among the features.

\begin{figure}[H]
\centering
\includegraphics[ width=0.95\linewidth]{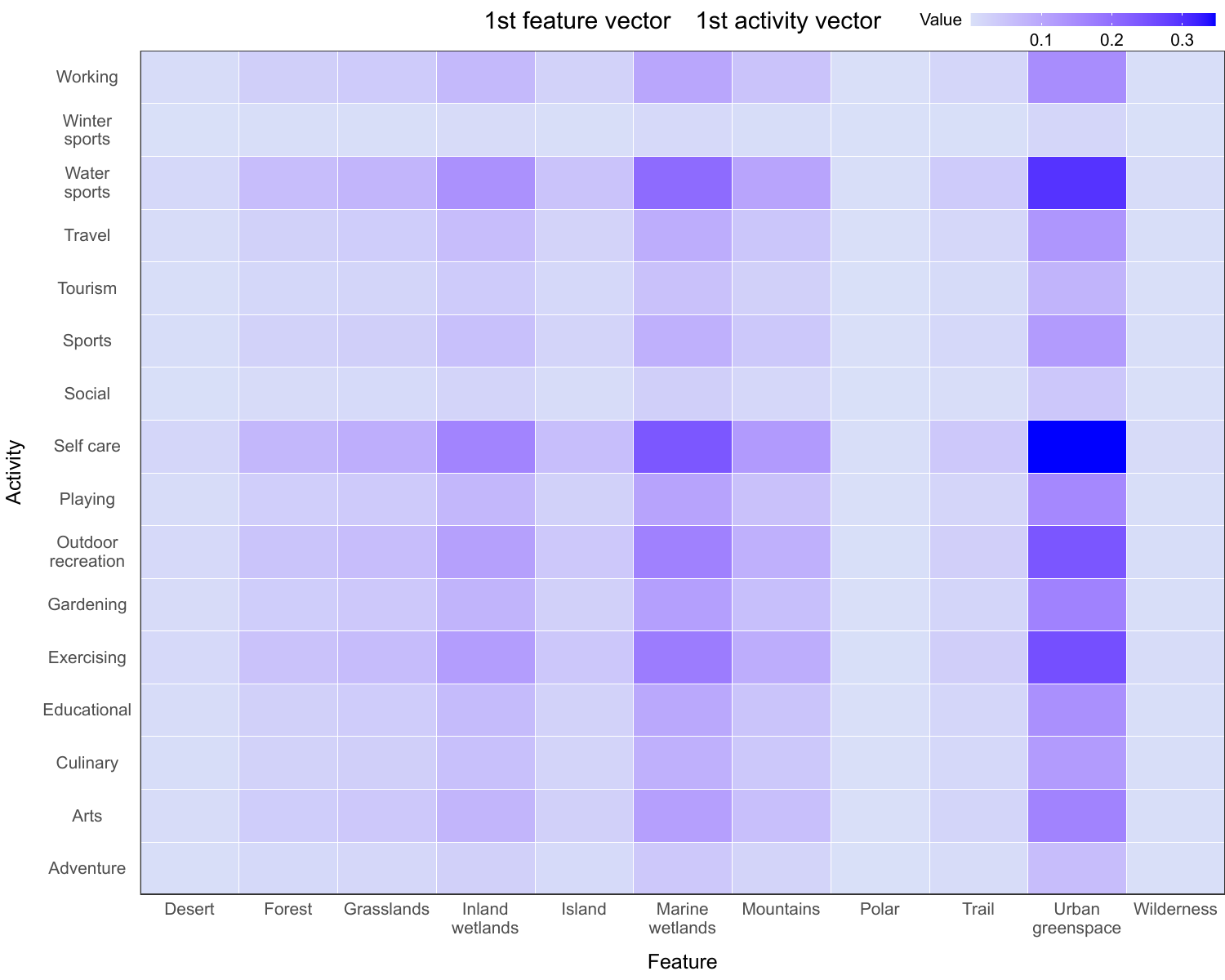}
\caption{Outer product matrix of the first feature (grouped) vector and first activity (grouped) vector.}
\label{fig:group_feat_act_product}
\end{figure}

At an annual scale, there were no noticeable temporal changes in the overall CES tweeting volume between the years (Fig. \ref{fig:time_series}, \ref{fig:wavelet_tweets}). However, noticeable changes in tweet volume were evident for some feature and activity classes, such as, urban greenspace features and social activities at the onset of the pandemic in March 2020 (Fig. \ref{fig:time_series}). We decomposed the \textbf{CES} tensor in its main components using a higher-order singular value decomposition (HOSVD, \cite{Kolda2009}). This decomposition highlighted that link level changes occurred for specific nature and activity classes. More particularly, the link between urban greenspace and self care activities was identified as a large contributor to the variance in the tensor along with the link between urban greenspace and outdoor recreational activities (Fig. \ref{fig:group_feat_act_product}), and between urban greenspace and water sports activities (primarily driven by swimming) (Appendix \ref{sec:outer_product}).

Moreover, an increased prevalence was observed between urban greenspace (garden, park, pool) and self-care activities (relaxing, meditating, thinking) during the first wave of the COVID-19 pandemic (Fig. \ref{fig:urban_activities_tweets_ts}a). This increased prevalence was temporary, and the link between urban greenspace and self care activities returned to its pre-crisis state the following year (2021), for which an increased prevalence was observed between urban greenspace and outdoor recreation activities (fishing, hiking, camping) (Fig. \ref{fig:urban_activities_tweets_ts}b). 

\begin{figure}[H]
\captionsetup[subfigure]{labelformat=empty}
\begin{subfigure}{0.5\textwidth}
  \centering
  \includegraphics[ width=\linewidth]{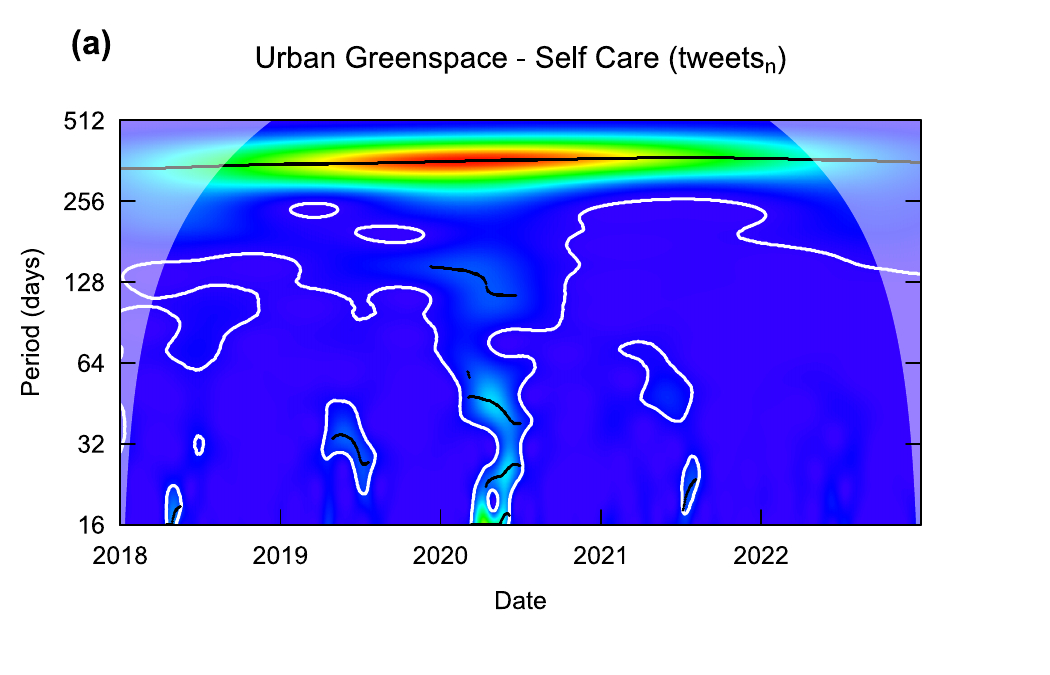}
\end{subfigure}
\begin{subfigure}{0.5\textwidth}
  \centering
  \includegraphics[ width=\linewidth]{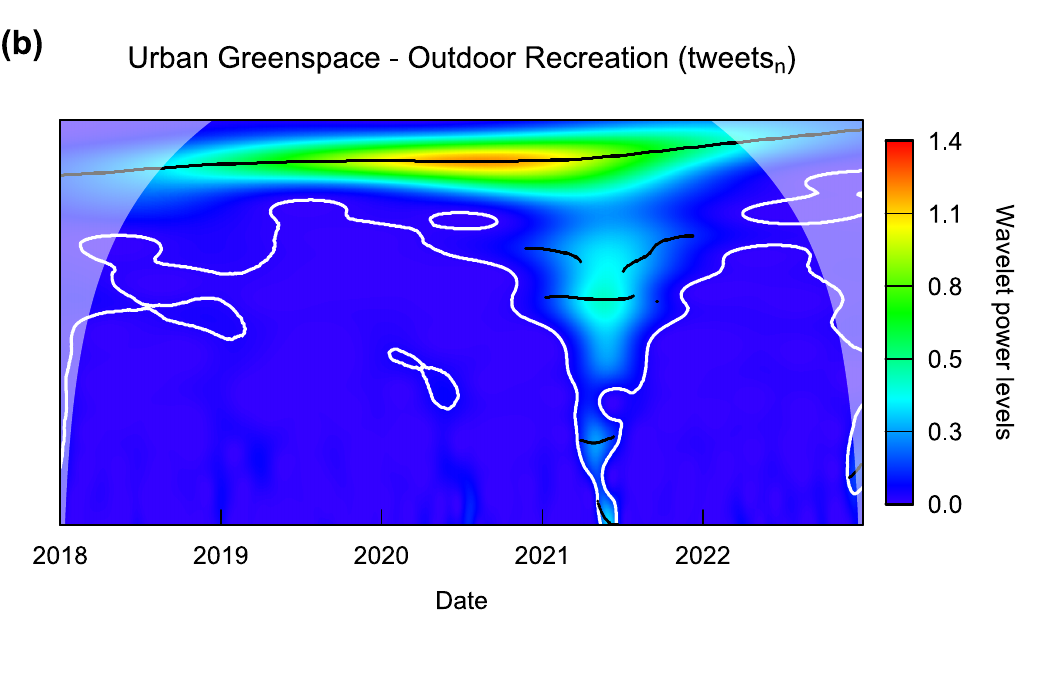}
\end{subfigure}
\caption{Wavelet analysis of the co-occurrence of \textbf{(a)} urban greenspace features and self care activities and \textbf{(b)} urban greenspace features and outdoor recreation activities in tweets from 2018-2022. White contour lines indicate significant time-period domains (95\% significance level). Black lines represent power ridges.}
\label{fig:greenspace_tweet_wavelets}
\end{figure}

The wavelet analysis of the two respective feature-activity pairs revealed an annual pattern (Fig. \ref{fig:greenspace_tweet_wavelets}). For the link between urban greenspace and self care a “short-term” disruption was identified from 13th of March to 31st of May 2020 (Fig. \ref{fig:greenspace_tweet_wavelets}a). This disruption extended a bit as we move through the 4 harmonics of the wavelet (at 16 days, about 30 days, about 46 days, and about 128 days. For the link between urban greenspace and outdoor recreation a “short-term” disruption was identified the following year from 17th of May to 15th of June 2021 (Fig. \ref{fig:greenspace_tweet_wavelets}b).

\begin{figure}[H]
\centering
\includegraphics[ width=0.95\linewidth]{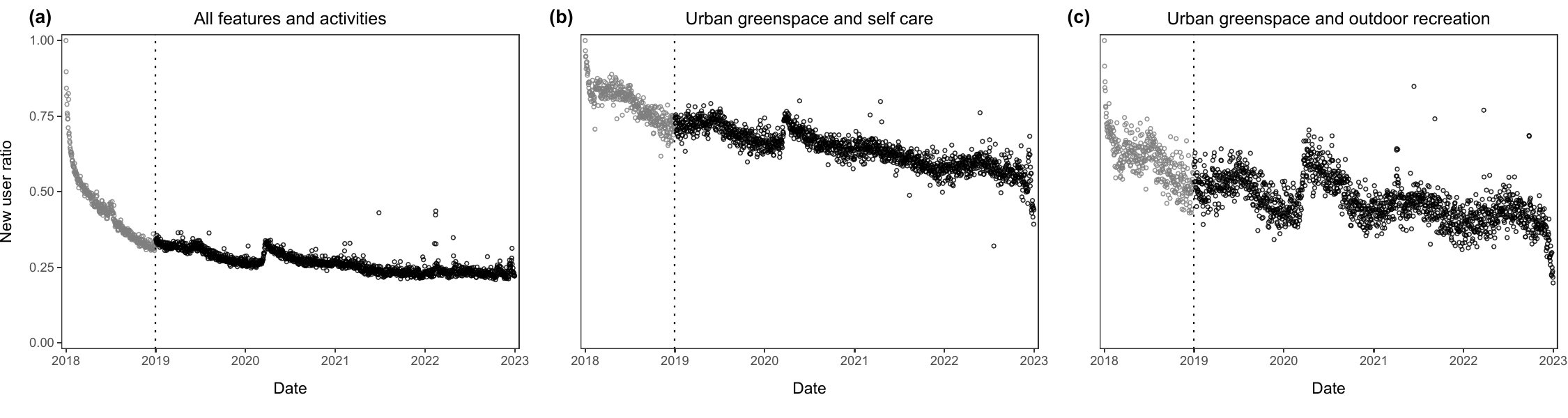}
\caption{Time series of the proportion of users tweeting about \textbf{(a)} CES features and activities, \textbf{(b)} urban greenspace features and self care activities, and \textbf{(c)} urban greenspace features and outdoor recreation activities for the first time.}
\label{fig:new_user_ratio}
\end{figure}

These tweet volume based network estimation could be biased by a few users tweeting a lot about particular activities or features. We, therefore, replicated the network estimation process (see \ref{sec:methods}) considering the number of daily users tweeting about feature-activity pairs rather that the total number of tweets for that day (see \ref{sec:methods}). This global CES user network, where links represent the number of users tweeting about feature-activity pairs, was comparable to the CES tweet network (Fig. \ref{fig:CES_tweet_network}, Fig. \ref{fig:CES_user_network}). Similar to the annual CES tweet networks, no noticeable changes were observed between the annual CES user networks (Fig. \ref{fig:CES_annual_user_network}). However, the daily proportion of users tweeting about CES features and activities for the first time revealed a large intake of new users tweeting about feature-activity pairs during the first wave of the COVID-19 pandemic, particularly in regard to urban greenspace and self care (Fig. \ref{fig:new_user_ratio}).

\begin{figure}[H]
\centering
\includegraphics[ width=0.95\linewidth]{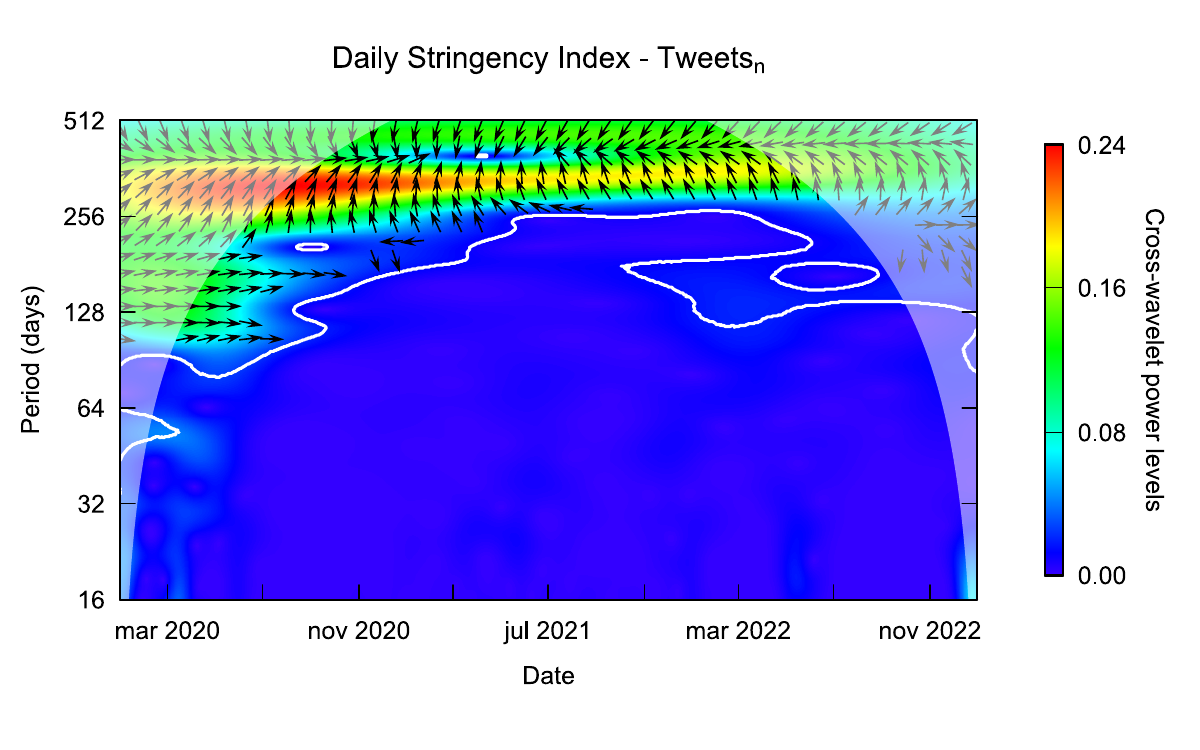}
\caption{Cross wavelet analysis between the COVID-19 stringency index (from \cite{OxCGRT_data}) and the number CES tweets from 2020-2022. The cross-wavelet spectrogram illustrates the time-frequency relationship between the two time series. The color intensity represents the strength of coherence between the two time series. White contour lines indicate significant time-period domains (95\% significance level). Arrows represent the phase relationship between the series. Right-pointing arrows indicate an in-phase relationship, suggesting synchronized variations between the stringency index and the number of tweets. Left-pointing arrows represent an out-of-phase relationship, suggesting opposite variations.}
\label{fig:SI_tweets_cross_wavelet}
\end{figure}

The cross wavelet analysis of COVID-19 mobility restrictions in relation to the ratio of new users tweeting about CES from 2020-2022 revealed an in-phase relationship between the two time series with an increased strength in coherency during the first wave of the pandemic (Fig. \ref{fig:SI_users_cross_wavelet}). Similarly, the cross wavelet analysis of mobility restrictions and the number of tweets also revealed an in-phase relationship between the two time series, but with the relationship being strongest around July 2020 (Fig. \ref{fig:SI_tweets_cross_wavelet}).

\section{Discussion} \label{sec:discuss}

We expected to observe large changes in the CES network topology given the large changes in global socioecological conditions during the COVID-19 pandemic associated with mobility restrictions across the globe \cite{OxCGRT_data}. This was not the case. There are several possibilities why we did not observe these expected changes. First, it is possible that our repertoires could not catch the activities or nature features for which conditions were most changed. Our multi-tier approach to aggregate features and activities in coarser categories might have not been the right scale of investigation. However, our replication of the analyses at the full repertoire scale did not change the outcome (Tab. \ref{tab:net_stats}). Second, the probability that the co-occurrence of activities and nature features was self-reported might have changed during the mobility restrictions, thus, masking topological changes. However, other studies using the same social media platform did not detect such changes \cite{Lusseau2023}. In addition, our analysis of the posting patterns of users (Fig. \ref{fig:CES_annual_user_network}) did not detect large changes in CES reporting. Therefore, we are left with explanations that are not associated with sampling biases. That is, self-reporting people maintained the same CES co-production profile. It is possible that the section of the population not on Twitter behaved differently, but there is no indication that would be the case, given the outcomes of small-scale studies using other sampling means \cite{Lusseau2023}. Given the mobility restrictions associated with the public health response to the pandemic, this would indicate that the vast majority of CES are produced locally and not by people traveling outside their usual 'habitat' to commonly seek CES. This has large implications for CES planning to maintain public wellbeing. 

Despite the overall structure of the global CES network being resilient to the COVID-19 perturbation, we observed an impulse on the link between self care activities and urban greenspace in response to the pandemic. This could be due to an increased need for self care activities during the pandemic and urban greenspace enabling CES to be produced locally. Thus, providing resilience for maintaining well-being during the pandemic \cite{Samuelsson2020}. Our user based analysis indicates a shift towards local CES production during the beginning of the pandemic, where there was a large increase in the proportion of users tweeting about specific CES features and activities for the first time. This could be due to an overall increase in users seeking CES, but the overall number of users followed the same annual temporal dynamics as the number of users. Thus, supporting that CES was produced locally. These findings suggest an overall need for CES and access to features providing CES in local communities. 
The link changes in the CES network were episodic and varied between the different waves of the pandemic. This echoes other large-scale studies showing that the way people used greenspace changed through the different pandemic waves \cite{Venter2020, Venter2021, Lusseau2023}. Overall the global CES network was resilient to a global perturbation of human behavior. 

\section{Methods} \label{sec:methods}
We outline here the data acquisition and pre-analytic procedures we use. For full replicability, we also have the code associated with these procedures available publicly as a Jupyter Notebook \cite{SoMe_CES_github}.

\subsection{Data collection}
\subsubsection{Compiling feature and activity lists}
A bottom-up approach was used to build the repertoire of CES human activities and CES nature features using Reddit, an assessed source of CES data deemed reliable for large-scale studies \cite{Fox2021} .
We queried all posts from the subreddit r/Outdoors over the past ten years and retrieved a list of all outdoor activities mentioned in these posts. The subreddit was selected because it has 4.7 million users and it is prescriptive in the way photos are to be titled. A list of activities was compiled using the UDPipe pipeline with the pre-trained English model to identify words of the gerund form \cite{UDPipe2019}.
This list was then filtered to ensure that the list only included outdoor activities performed by humans, resulting in a list of 186 human activities retrieved from 92,801 posts \cite{SoMe_CES_github}. A similar approach was used to compile a list of nature features. Posts were queried from the subreddit r/EarthPorn over the past ten years. This subreddit was selected because it is now the largest online community of nature photographers with 23.3 million users. It is also very prescriptive in the way photos have to be titled, easing the process of identifying features posted. Nature features were then found and extracted using the UDPipe pipeline on the post titles with the pre-trained English model to find nouns with a root dependency in post texts \cite{UDPipe2019, SoMe_CES_github}. This list of nouns was then used to compile a complete list of nature features, resulting in a list of 39 main nature features retrieved from 589,405 posts.

\subsubsection{Twitter sampling}
These lists of nature features and human activities were then used to sample Twitter (using the Academic Research API) for English tweets mentioning one feature and one activity over a five year period (2018-2022) \cite{SoMe_CES_github}. Approximately 68 million tweets were retrieved (Table \ref{tab:n_tweets}). However, specific nature features and human activities were repeatedly also used in other contexts, e.g. nature features used as part of sports team names or school names. Such misclassifications were removed using context information provided by Twitter for each tweet \cite{SoMe_CES_github, twitter_context_annotations} , resulting in a dataset of 41.7 million tweets (Table \ref{tab:n_tweets}).

\begin{table}[h]
\begin{center}
\begin{minipage}{\textwidth}
\caption{Number of tweets table}\label{tab:n_tweets}
\begin{tabular*}{\textwidth}{@{\extracolsep{\fill}}lccc@{\extracolsep{\fill}}}
\toprule%
     & \multicolumn{3}{c}{\textbf{Tweets (n)}} \\
\textbf{Year} & \multicolumn{1}{c}{Queried} & \multicolumn{1}{c}{After inital cleaning} & \multicolumn{1}{c}{After final cleaning} \\ \midrule
2018 & 13,190,894 & 8,288,414 & 7,884,298 \\
2019 & 13,630,403 & 8,817,854 & 8,550,859 \\
2020 & 12,727,849 & 8,458,657 & 8,211,259 \\
2021 & 13,192,365 & 8,746,318 & 8,491,994 \\
2022 & 14,697,965 & 8,881,359 & 8,577,815 \\ 
\botrule
\end{tabular*}
\end{minipage}
\end{center}
\end{table}

\subsection{Analysis}
\subsubsection{Network construction and analysis}
From these tweets, we constructed a bipartite network, where each link was weighted by the number of times nature features and human activities co-occurred in tweets over the five year period (2018-2022). Due to the large number of features and activities, a second network was constructed, where features and activities were pooled into larger classes, i.e., features were grouped into 11 nature classes and activities were grouped into 16 activity classes. Nature features were grouped by nature type based on physical characteristics. We used the CaLA (Catalogue of Leisure Activities) \cite{activity_catalogue} as a tool for determining activity class \cite{SoMe_CES_github}. The resulting CES networks (full and grouped) were also estimated annually for 2018 to 2022. This period covered tweets reporting human-nature interactions before, during, and after global mobility public health interventions were in place to combat the COVID-19 pandemic. Thus, enabling us to compare the network topological state  changes in response to this global perturbation.

These bipartite networks, networks with links between two different types of nodes, were structurally compared to determine if they represent different states and compare the role of features and activities in the network structure. We used global measures of network asymmetry, modularity, nestedness, connectance, and interaction asymmetry were also calculated as prospective dimensions along which network states could be defined. Moreover, node level measures of interaction asymmetry and nested rank were used to determine the most central nodes in the network, thus, identifying the most important nature features and human activities in the annual networks.

\subsubsection{Temporal dynamics of CES co-production}
To assess the temporal dynamics of the overall CES co-production reported on Twitter, a time series was created with the number of daily tweets mentioning human activities and nature features. Time series were also created for subsets of tweets mentioning a specific feature or activity class.

\subsubsection{Higher-Order Singular Value Decomposition}
As we have the same nodes for each year, we were able to also represent the CES network as a CES interaction three-dimensional tensor (activities x features x day). We used a higher order singular value decomposition to decompose the tensor. The outer product of the first vectors of the resulting factor matrices ($U_{1-3}$) were computed. Thus, providing information about the relationships and dependencies in the original tensor that are captured by the HOSVD.

\subsubsection{Wavelet analysis}
We used a spectral decomposition approach to assess the primary temporal changes in this CES network tensor representation. A wavelet analysis was used to capture temporal dynamics of the overall CES tweet volume at different frequencies. We also assessed the time-frequency dynamics of selected links (feature-activity pairs) that were identified as important in the HOSVD analysis.

\subsubsection{User based network analysis}
We also constructed both global and annual bipartite networks (full and grouped) based on the number of users mentioning nature features and human activities on Twitter for the five year period (2018-2022), i.e. links were weighted by the number of users instead of the number of tweets. The topological state changes of these user based networks were compared to that of the tweet based networks. 

\subsubsection{CES user turnover}
We calculated the daily proportion of new CES users, i.e. users that had not previously tweeted about CES features and activities. The daily proportion of new users was also calculated for the feature-activity pairs identified in the tweet based analysis. We created time series' of the new user ratios to identify the temporal dynamics of the user turnover. We also conducted a wavelet analysis of the new user ratios to understand the temporal dynamics of the CES user turnover at different frequencies.

\subsubsection{Coherency analysis between COVID-19 mobility restrictions and CES reported on Twitter} 
To assess the relationship between CES reported on Twitter and mobility restrictions associated with the COVID-19 pandemic we compared the number of CES tweets and proportion of new CES users in relation to the mobility restrictions associated with the pandemic. To estimate mobility restrictions we used the stringency index from the Oxford COVID-19 Government Response Tracker \cite{OxCGRT_data} and calculated the median stringency of English speaking countries. A cross-wavelet analysis was then used to estimate the coherency between the stringency index and the number of CES tweets. We did the same analysis for the ratio of new CES users.

Cross wavelet analysis between the COVID-19 stringency index (from \cite{OxCGRT_data}) and the ratio of new users tweeting about CES from 2020-2022. The cross-wavelet spectrogram illustrates the time-frequency relationship between the two time series. The color intensity represents the strength of coherence between the two time series. White contour lines indicate significant time-period domains (95\% significance level). Arrows represent the phase relationship between the series. Right-pointing arrows indicate an in-phase relationship, suggesting synchronized variations between the stringency index and the new user ratio. Left-pointing arrows represent an out-of-phase relationship, suggesting opposite variations.

\backmatter








\noindent

\begin{appendices}

\section{Annual CES network}\label{sec:yearly_tweet_network_appendix}
\begin{figure}[H]
\centering
\includegraphics[ width=0.99\linewidth]{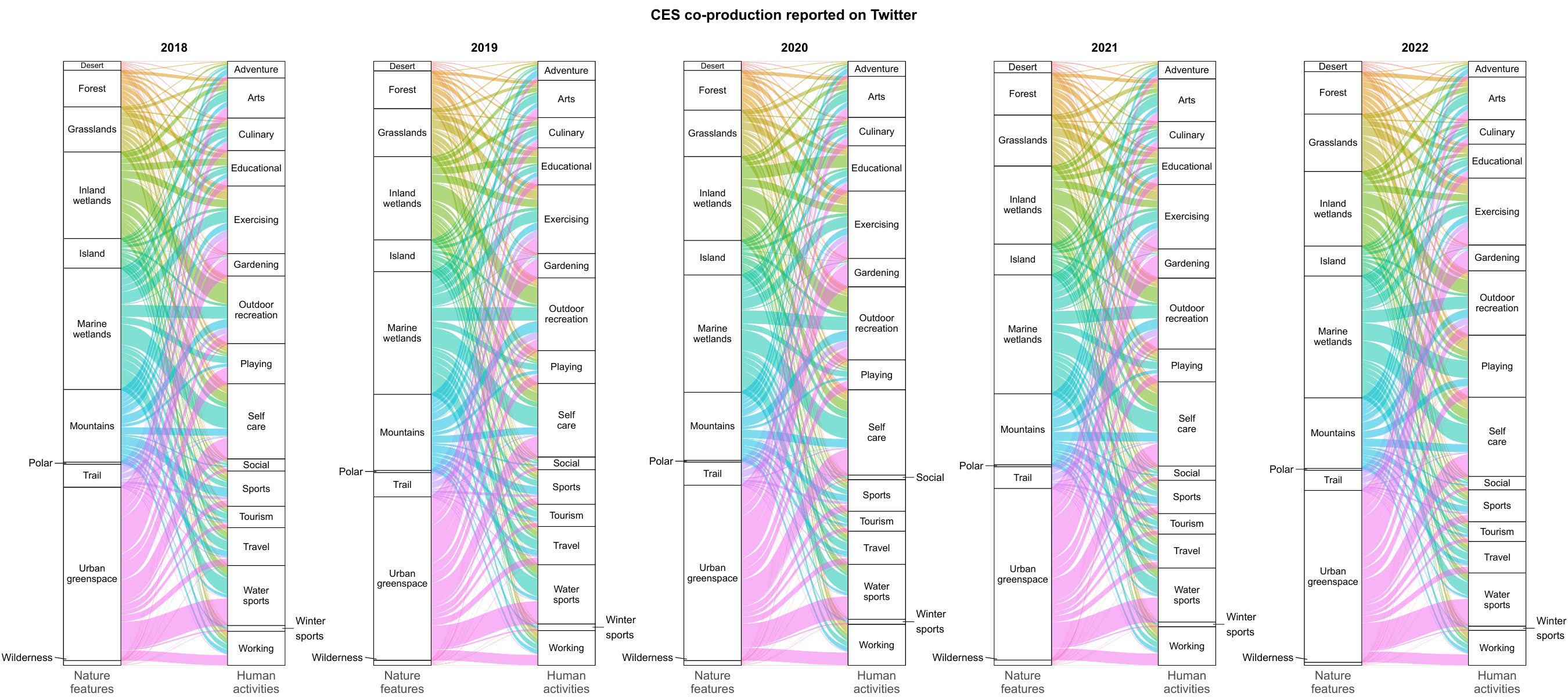}
\caption{Annual CES networks of tweets mentioning human-nature interactions on Twitter from 2018-2022 (41.7 million tweets). Links are weighted by the number of tweets including a given nature feature (grouped by nature class) and human activity (grouped by activity class).}
\label{fig:CES_annual_tweet_network}
\end{figure}

\section{Network level statistics}\label{sec:network_stats_appendix}

\begin{table}[h]
\caption{}
\label{tab:net_stats}
\begin{tabular}{lcccccc}
\toprule
\multicolumn{1}{r}{\textbf{}}  & \textbf{2018} & \textbf{2019} & \textbf{2020} & \textbf{2021} & \textbf{2022} & \textbf{All years} \\ \cmidrule(l){2-7} 
                               & \multicolumn{6}{c}{Grouped}                                                                        \\ \cmidrule(l){2-7} 
\textbf{Web asymmetry}         & 0.185         & 0.185         & 0.185         & 0.185         & 0.185         & 0.185              \\
\textbf{Modularity}            & 0.128         & 0.117         & 0.121         & 0.125         & 0.120         & 0.121              \\
\textbf{Weighted nestedness}   & 0.0901        & 0.0752        & 0.0909        & 0.0708        & 0.0471        & 0.0748             \\
\textbf{Interaction asymmetry} & 0.0284        & 0.0284        & 0.0284        & 0.0284        & 0.0284        & 0.0284             \\
\textbf{Weighted connectance}  & 0.354         & 0.361         & 0.347         & 0.358         & 0.355         & 0.358                                             \\ \cmidrule(l){2-7} 
                               & \multicolumn{6}{c}{Full}                                                                           \\ \cmidrule(l){2-7} 
\textbf{Web asymmetry}         & 0.647         & 0.653         & 0.653         & 0.653         & 0.647         & 0.653              \\
\textbf{Modularity}            & 0.219         & 0.207         & 0.216         & 0.207         & 0.201         & 0.203              \\
\textbf{Weighted nestedness}   & 0.555         & 0.551         & 0.528         & 0.562         & 0.578         & 0.451              \\
\textbf{Interaction asymmetry} & 0.0211        & 0.0211        & 0.0212        & 0.0213        & 0.0214        & 0.0207             \\
\textbf{Weighted connectance}  & 0.142         & 0.150         & 0.143         & 0.149         & 0.137         & 0.147              \\ \bottomrule
\end{tabular}
\end{table}

\newpage

\section{Node level statistics}\label{sec:node_level_stats}
\begin{table}[h]
\caption{}
\label{tab:node_stats}
\addtolength{\tabcolsep}{-3pt}
\footnotesize
\begin{tabular}{@{}lcccccccccccc@{}}
\toprule
 & \multicolumn{6}{c}{\textbf{Interaction asymmetry}} & \multicolumn{6}{c}{\textbf{Nested rank}} \\ \midrule
\textbf{} & \textbf{2018} & \textbf{2019} & \textbf{2020} & \textbf{2021} & \textbf{2022} & \textbf{All} & \textbf{2018} & \textbf{2019} & \textbf{2020} & \textbf{2021} & \textbf{2022} & \textbf{All} \\ \midrule
\multicolumn{13}{c}{\textbf{Activities}} \\ \midrule
\textbf{Adventure} & -0.0612 & -0.0563 & -0.0627 & -0.0618 & -0.0634 & -0.0610 & 0.870 & 0.870 & 0.870 & 0.870 & 0.870 & 0.870 \\
\textbf{Arts} & -0.0280 & -0.0333 & -0.0238 & -0.0222 & -0.0196 & -0.0257 & 0.330 & 0.330 & 0.400 & 0.270 & 0.330 & 0.270 \\
\textbf{Culinary} & -0.0371 & -0.0415 & -0.0471 & -0.0501 & -0.0498 & -0.0453 & 0.670 & 0.670 & 0.730 & 0.730 & 0.730 & 0.670 \\
\textbf{Educational} & -0.0310 & -0.0278 & -0.0128 & -0.0226 & -0.0242 & -0.0237 & 0.470 & 0.400 & 0.270 & 0.400 & 0.470 & 0.400 \\
\textbf{Exercising} & 0.0358 & 0.0283 & 0.0259 & 0.0201 & 0.0273 & 0.0272 & 0.067 & 0.130 & 0.130 & 0.130 & 0.067 & 0.130 \\
\textbf{Gardening} & -0.0733 & -0.0707 & -0.0685 & -0.0680 & -0.0707 & -0.0702 & 0.730 & 0.730 & 0.670 & 0.670 & 0.670 & 0.730 \\
\textbf{\begin{tabular}[c]{@{}l@{}}Outdoor \\ recreation\end{tabular}} & 0.0526 & 0.0685 & 0.0667 & 0.0774 & 0.0518 & 0.0647 & 0.130 & 0.067 & 0.067 & 0.067 & 0.130 & 0.067 \\
\textbf{Playing} & -0.0365 & -0.0420 & -0.0448 & -0.0411 & 0.000652 & -0.0336 & 0.270 & 0.600 & 0.600 & 0.600 & 0.200 & 0.330 \\
\textbf{Self care} & 0.0249 & 0.0172 & 0.0388 & 0.0409 & 0.0310 & 0.0302 & 0.000 & 0.000 & 0.000 & 0.000 & 0.000 & 0.000 \\
\textbf{Social} & -0.0707 & -0.0702 & -0.0837 & -0.0658 & -0.0673 & -0.0716 & 0.930 & 0.930 & 1.00 & 0.930 & 0.930 & 0.930 \\
\textbf{Sports} & -0.0293 & -0.0331 & -0.0354 & -0.0357 & -0.0354 & -0.0339 & 0.530 & 0.530 & 0.530 & 0.530 & 0.530 & 0.600 \\
\textbf{Tourism} & -0.0482 & -0.0406 & -0.0509 & -0.0524 & -0.0500 & -0.0482 & 0.800 & 0.800 & 0.800 & 0.800 & 0.800 & 0.800 \\
\textbf{Travel} & -0.0141 & -0.0189 & -0.0271 & -0.0289 & -0.0325 & -0.0241 & 0.400 & 0.270 & 0.470 & 0.470 & 0.600 & 0.530 \\
\textbf{\begin{tabular}[c]{@{}l@{}}Water \\ sports\end{tabular}} & -0.0272 & -0.0248 & -0.0322 & -0.0358 & -0.0380 & -0.0314 & 0.200 & 0.200 & 0.200 & 0.200 & 0.270 & 0.200 \\
\textbf{\begin{tabular}[c]{@{}l@{}}Winter \\ sports\end{tabular}} & -0.0737 & -0.0732 & -0.0754 & -0.0785 & -0.0778 & -0.0757 & 1.000 & 1.000 & 0.930 & 1.000 & 1.000 & 1.000 \\
\textbf{Working} & -0.0374 & -0.0363 & -0.0216 & -0.0301 & -0.0365 & -0.0323 & 0.600 & 0.470 & 0.330 & 0.330 & 0.400 & 0.470 \\ \midrule
\multicolumn{13}{c}{\textbf{Features}} \\ \midrule
\textbf{Desert} & \multicolumn{1}{r}{-0.0467} & \multicolumn{1}{r}{-0.046} & \multicolumn{1}{r}{-0.0465} & \multicolumn{1}{r}{-0.0427} & \multicolumn{1}{r}{-0.0443} & \multicolumn{1}{r}{-0.0452} & \multicolumn{1}{r}{0.8} & \multicolumn{1}{r}{0.8} & \multicolumn{1}{r}{0.8} & \multicolumn{1}{r}{0.8} & \multicolumn{1}{r}{0.8} & \multicolumn{1}{r}{0.8} \\
\textbf{Forest} & \multicolumn{1}{r}{-0.00244} & \multicolumn{1}{r}{-0.000591} & \multicolumn{1}{r}{0.00185} & \multicolumn{1}{r}{0.00611} & \multicolumn{1}{r}{0.00602} & \multicolumn{1}{r}{0.00195} & \multicolumn{1}{r}{0.5} & \multicolumn{1}{r}{0.5} & \multicolumn{1}{r}{0.5} & \multicolumn{1}{r}{0.5} & \multicolumn{1}{r}{0.5} & \multicolumn{1}{r}{0.5} \\
\textbf{Grasslands} & \multicolumn{1}{r}{0.0113} & \multicolumn{1}{r}{0.0168} & \multicolumn{1}{r}{0.0145} & \multicolumn{1}{r}{0.0202} & \multicolumn{1}{r}{0.0271} & \multicolumn{1}{r}{0.0178} & \multicolumn{1}{r}{0.4} & \multicolumn{1}{r}{0.4} & \multicolumn{1}{r}{0.4} & \multicolumn{1}{r}{0.4} & \multicolumn{1}{r}{0.4} & \multicolumn{1}{r}{0.4} \\
\textbf{Inland wetlands} & \multicolumn{1}{r}{0.0703} & \multicolumn{1}{r}{0.0673} & \multicolumn{1}{r}{0.0651} & \multicolumn{1}{r}{0.0587} & \multicolumn{1}{r}{0.0592} & \multicolumn{1}{r}{0.0647} & \multicolumn{1}{r}{0.2} & \multicolumn{1}{r}{0.2} & \multicolumn{1}{r}{0.2} & \multicolumn{1}{r}{0.2} & \multicolumn{1}{r}{0.2} & \multicolumn{1}{r}{0.2} \\
\textbf{Island} & \multicolumn{1}{r}{-0.0146} & \multicolumn{1}{r}{-0.0109} & \multicolumn{1}{r}{-0.00647} & \multicolumn{1}{r}{-0.0125} & \multicolumn{1}{r}{-0.0134} & \multicolumn{1}{r}{-0.0115} & \multicolumn{1}{r}{0.6} & \multicolumn{1}{r}{0.6} & \multicolumn{1}{r}{0.6} & \multicolumn{1}{r}{0.6} & \multicolumn{1}{r}{0.6} & \multicolumn{1}{r}{0.6} \\
\textbf{Marine wetlands} & \multicolumn{1}{r}{0.1230} & \multicolumn{1}{r}{0.125} & \multicolumn{1}{r}{0.120} & \multicolumn{1}{r}{0.123} & \multicolumn{1}{r}{0.123} & \multicolumn{1}{r}{0.123} & \multicolumn{1}{r}{0.1} & \multicolumn{1}{r}{0.1} & \multicolumn{1}{r}{0.1} & \multicolumn{1}{r}{0.1} & \multicolumn{1}{r}{0.1} & \multicolumn{1}{r}{0.1} \\
\textbf{Mountains} & \multicolumn{1}{r}{0.0787} & \multicolumn{1}{r}{0.0824} & \multicolumn{1}{r}{0.0713} & \multicolumn{1}{r}{0.0718} & \multicolumn{1}{r}{0.0737} & \multicolumn{1}{r}{0.0761} & \multicolumn{1}{r}{0.3} & \multicolumn{1}{r}{0.3} & \multicolumn{1}{r}{0.3} & \multicolumn{1}{r}{0.3} & \multicolumn{1}{r}{0.3} & \multicolumn{1}{r}{0.3} \\
\textbf{Polar} & \multicolumn{1}{r}{-0.0584} & \multicolumn{1}{r}{-0.0581} & \multicolumn{1}{r}{-0.0592} & \multicolumn{1}{r}{-0.0590} & \multicolumn{1}{r}{-0.0588} & \multicolumn{1}{r}{-0.0586} & \multicolumn{1}{r}{1.0} & \multicolumn{1}{r}{1.0} & \multicolumn{1}{r}{1.0} & \multicolumn{1}{r}{1.0} & \multicolumn{1}{r}{1.0} & \multicolumn{1}{r}{1.0} \\
\textbf{Trail} & \multicolumn{1}{r}{-0.0278} & \multicolumn{1}{r}{-0.0265} & \multicolumn{1}{r}{-0.0275} & \multicolumn{1}{r}{-0.0286} & \multicolumn{1}{r}{-0.0309} & \multicolumn{1}{r}{-0.0282} & \multicolumn{1}{r}{0.7} & \multicolumn{1}{r}{0.7} & \multicolumn{1}{r}{0.7} & \multicolumn{1}{r}{0.7} & \multicolumn{1}{r}{0.7} & \multicolumn{1}{r}{0.7} \\
\textbf{Urban greenspace} & \multicolumn{1}{r}{0.2320} & \multicolumn{1}{r}{0.215} & \multicolumn{1}{r}{0.231} & \multicolumn{1}{r}{0.228} & \multicolumn{1}{r}{0.226} & \multicolumn{1}{r}{0.225} & \multicolumn{1}{r}{0.0} & \multicolumn{1}{r}{0.0} & \multicolumn{1}{r}{0.0} & \multicolumn{1}{r}{0.0} & \multicolumn{1}{r}{0.0} & \multicolumn{1}{r}{0.0} \\
\textbf{Wilderness} & \multicolumn{1}{r}{-0.0525} & \multicolumn{1}{r}{-0.0525} & \multicolumn{1}{r}{-0.0516} & \multicolumn{1}{r}{-0.0525} & \multicolumn{1}{r}{-0.0552} & \multicolumn{1}{r}{-0.0528} & \multicolumn{1}{r}{0.9} & \multicolumn{1}{r}{0.9} & \multicolumn{1}{r}{0.9} & \multicolumn{1}{r}{0.9} & \multicolumn{1}{r}{0.9} & \multicolumn{1}{r}{0.9} \\ \bottomrule
\end{tabular}
\end{table}

\begin{figure}[H]
\centering
\includegraphics[ width=0.99\linewidth]{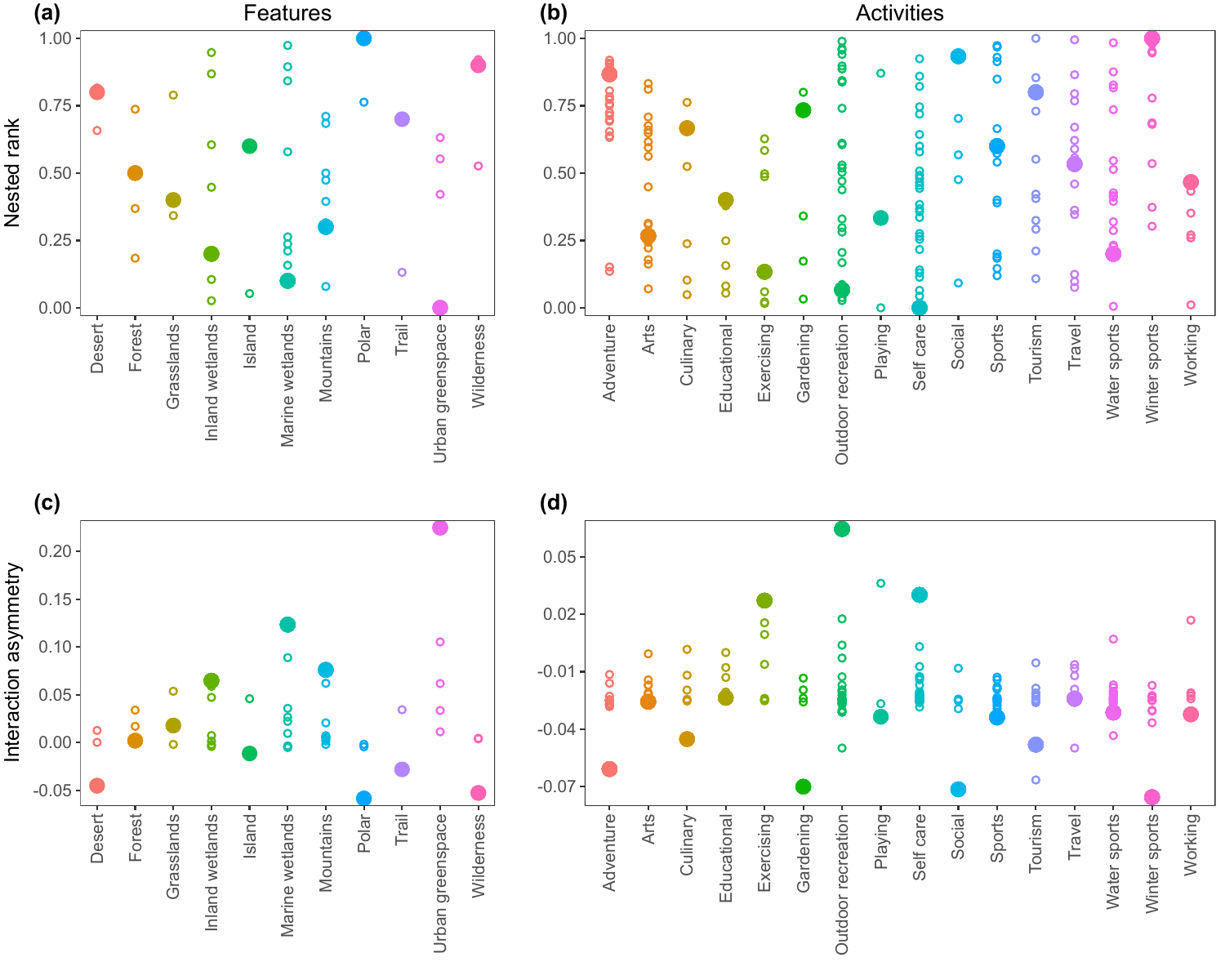}
\caption{}
\label{fig:node_level_stats}
\end{figure}

\section{Time series}\label{sec:time_series}

\begin{figure}[H]
\centering
\includegraphics[ width=0.95\linewidth]{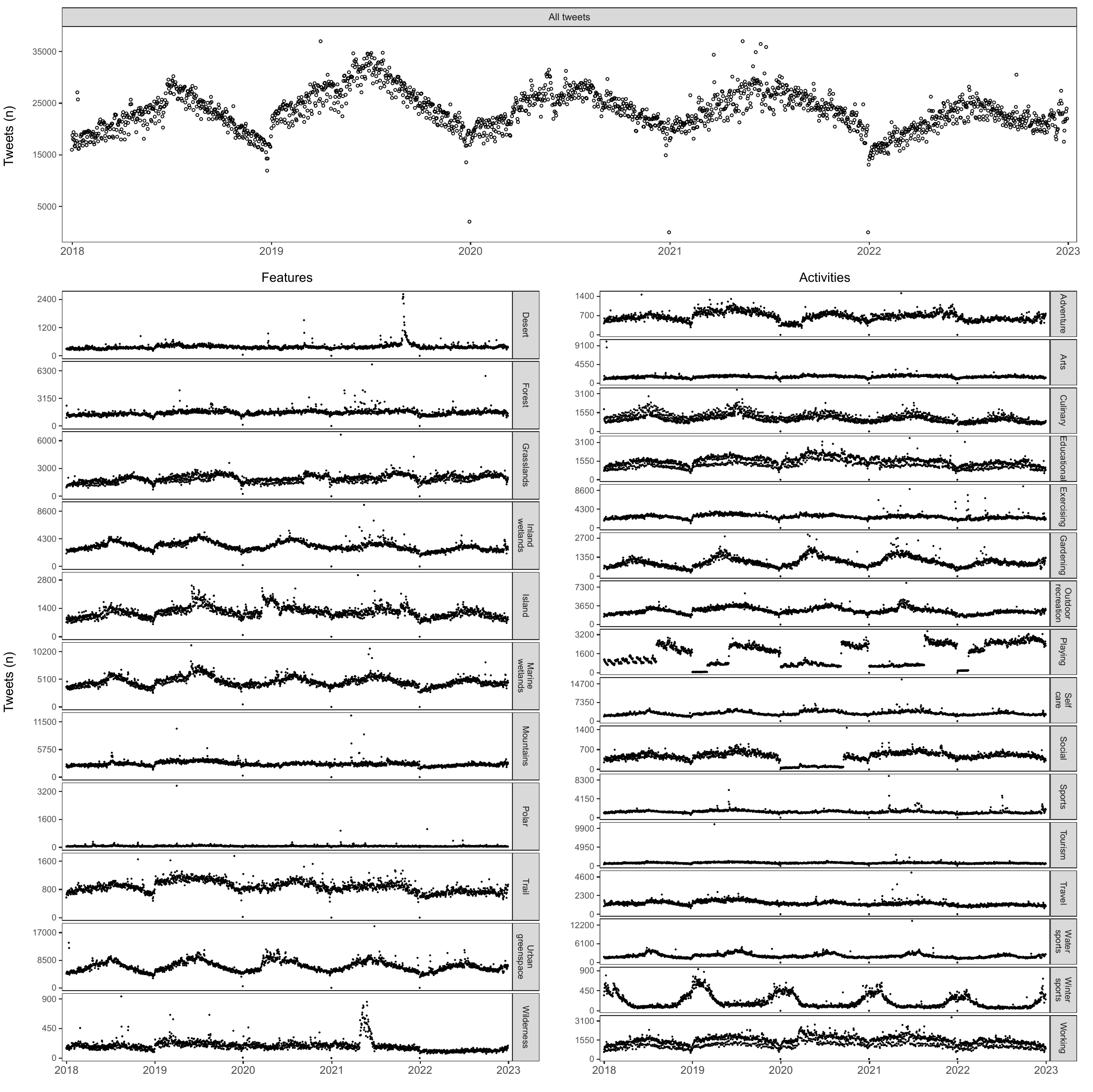}
\caption{Time series of the co-occurrence of nature features and human activities in tweets from 2018-2022. The top panel shows the time series of the overall number of CES tweets. The left panel shows the number of CES tweets for each nature feature class and the right panel shows the number of CES tweets for each human activity class.}
\label{fig:time_series}
\end{figure}

\begin{figure}[H]
\centering
\includegraphics[width=0.75\linewidth]{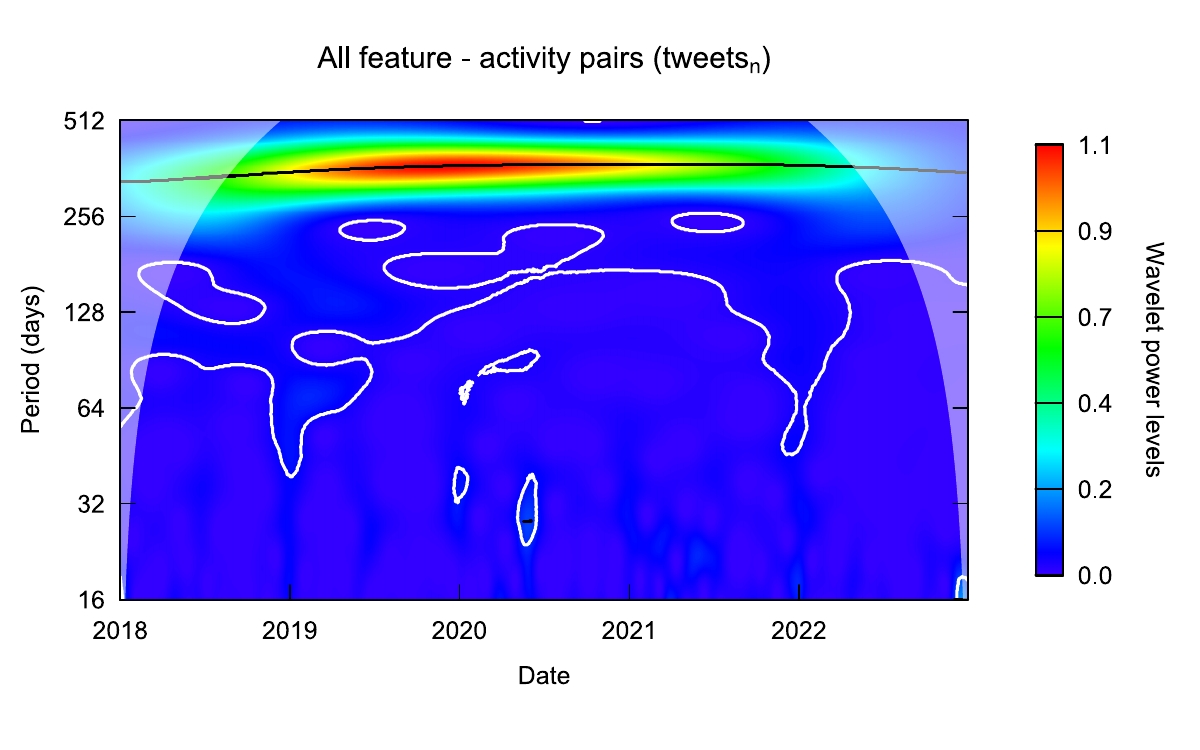}
\caption{Wavelet analysis of the co-occurrence of nature features and human activities in tweets from 2018-2022. White contour lines indicate significant time-period domains (95\% significance level). Black lines represent power ridges.}
\label{fig:wavelet_tweets}
\end{figure}

\section{Outer product}\label{sec:outer_product}
\begin{figure}[H]
\centering
\includegraphics[ width=0.95\linewidth]{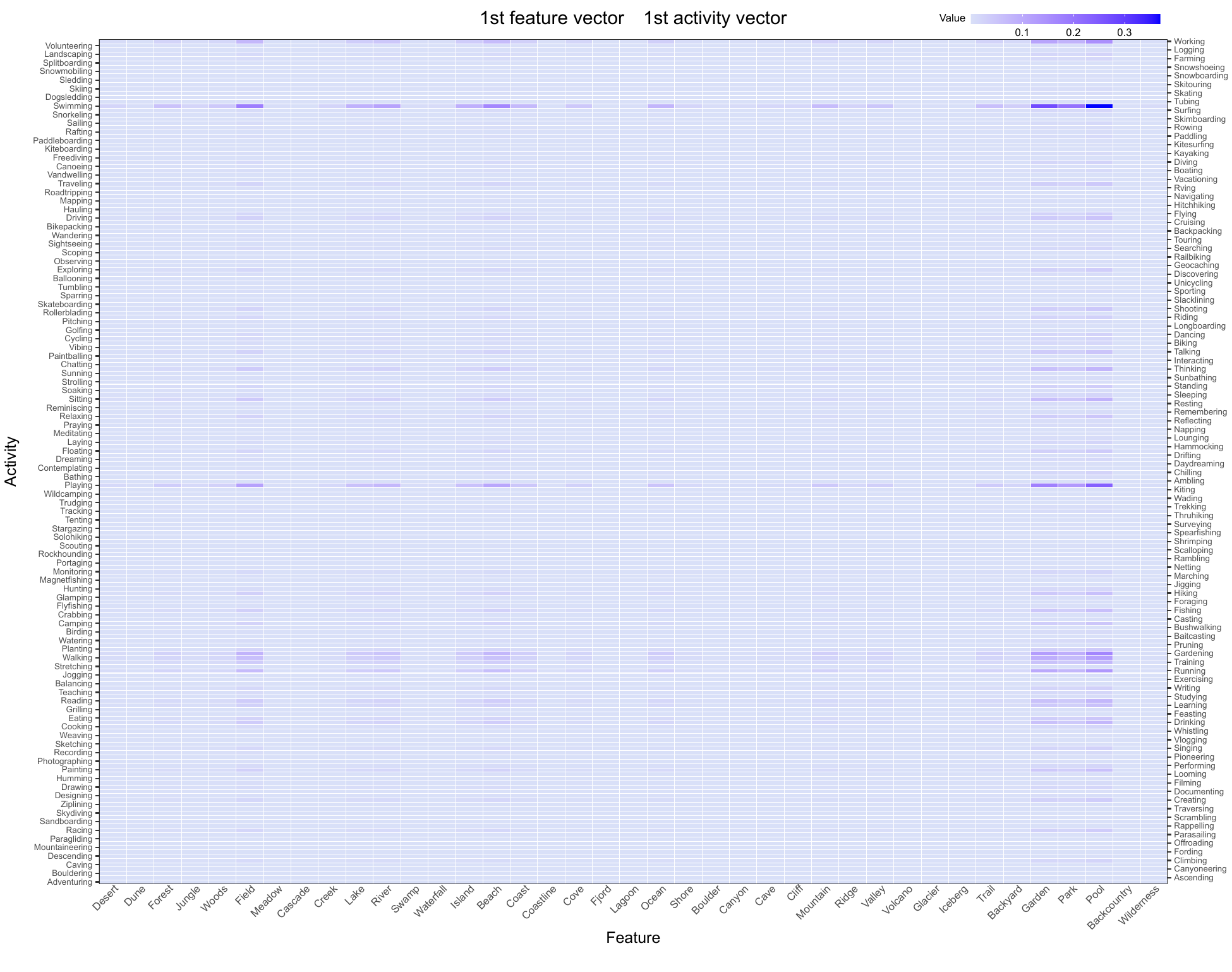}
\caption{Outer product of the first feature (full) vector and first activity (full) vector.}
\label{fig:full_feat_act_product}
\end{figure}

\begin{figure}[H]
\centering
\includegraphics[ width=0.95\linewidth]{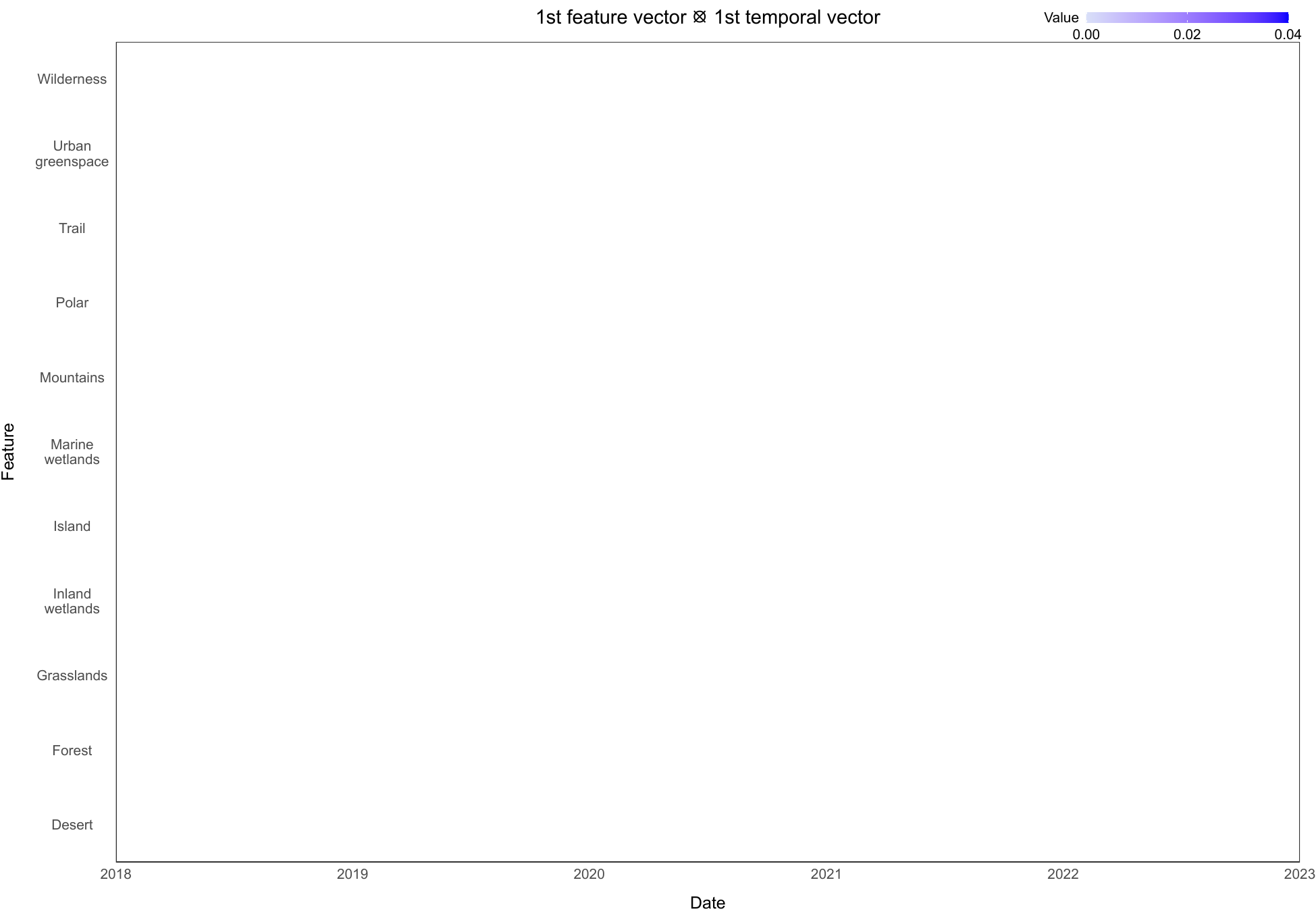}
\caption{Outer product of the first feature (grouped) vector and first temporal vector.}
\label{fig:group_feat_time_product}
\end{figure}

\begin{figure}[H]
\centering
\includegraphics[ width=0.95\linewidth]{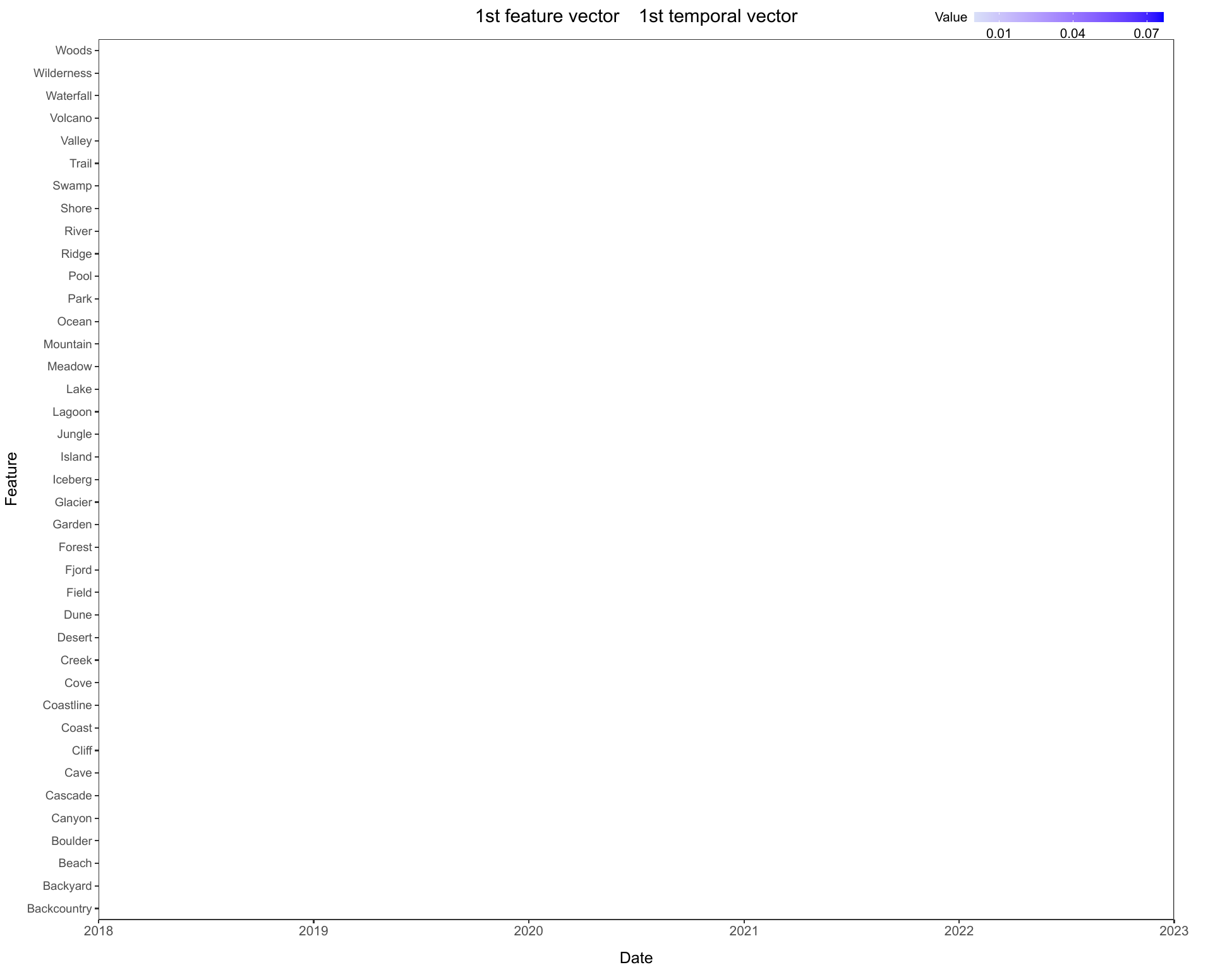}
\caption{Outer product of the first feature (full) vector and first temporal vector.}
\label{fig:full_feat_time_product}
\end{figure}

\begin{figure}[H]
\centering
\includegraphics[ width=0.95\linewidth]{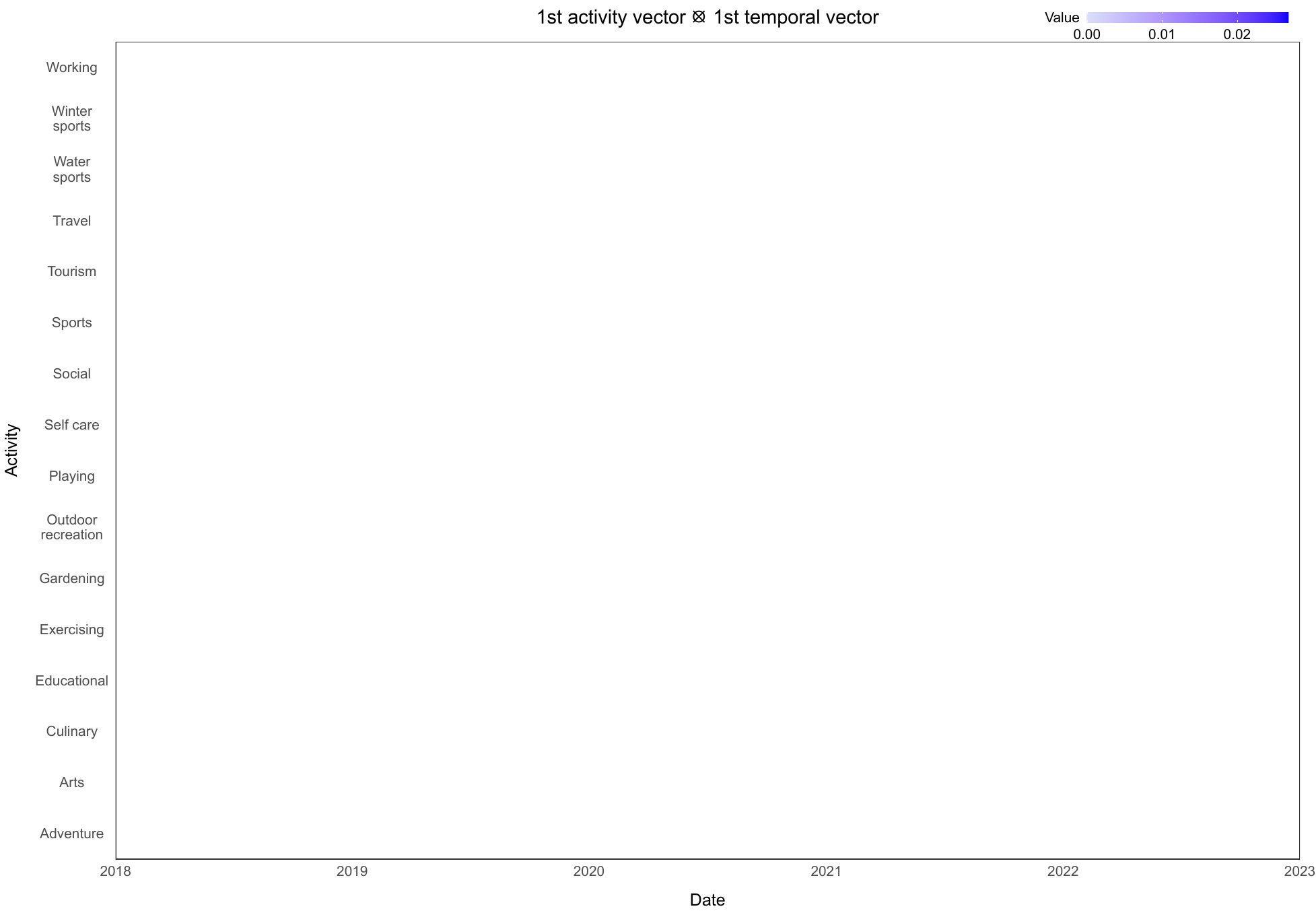}
\caption{Outer product of the first activity (grouped) vector and first temporal vector.}
\label{fig:group_act_time_product}
\end{figure}

\begin{figure}[H]
\centering
\includegraphics[ width=0.95\linewidth]{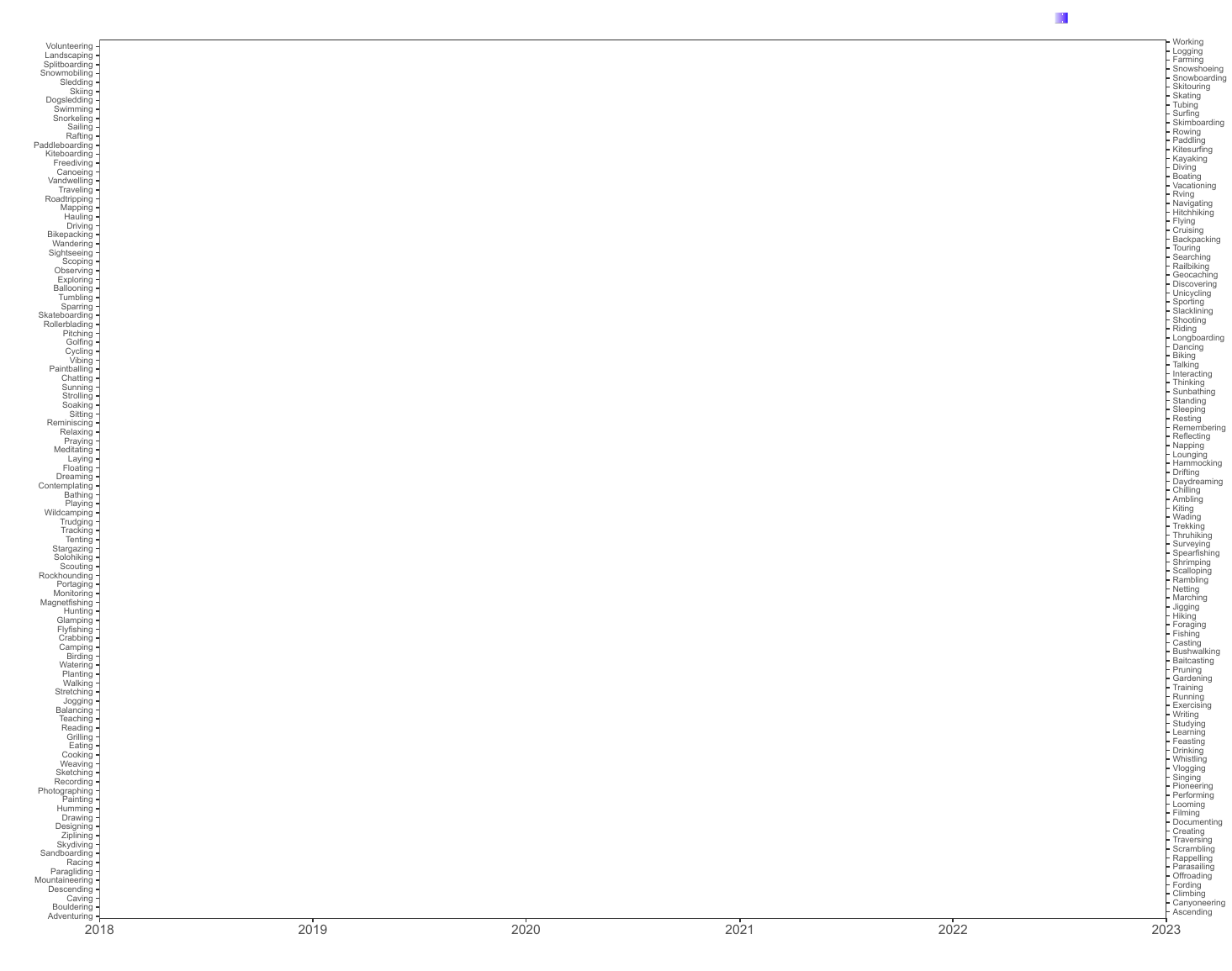}
\caption{Outer product of the first activity (full) vector and first temporal vector.}
\label{fig:full_act_time_product}
\end{figure}

\begin{figure}[H]
\centering
\includegraphics[ width=0.95\linewidth]{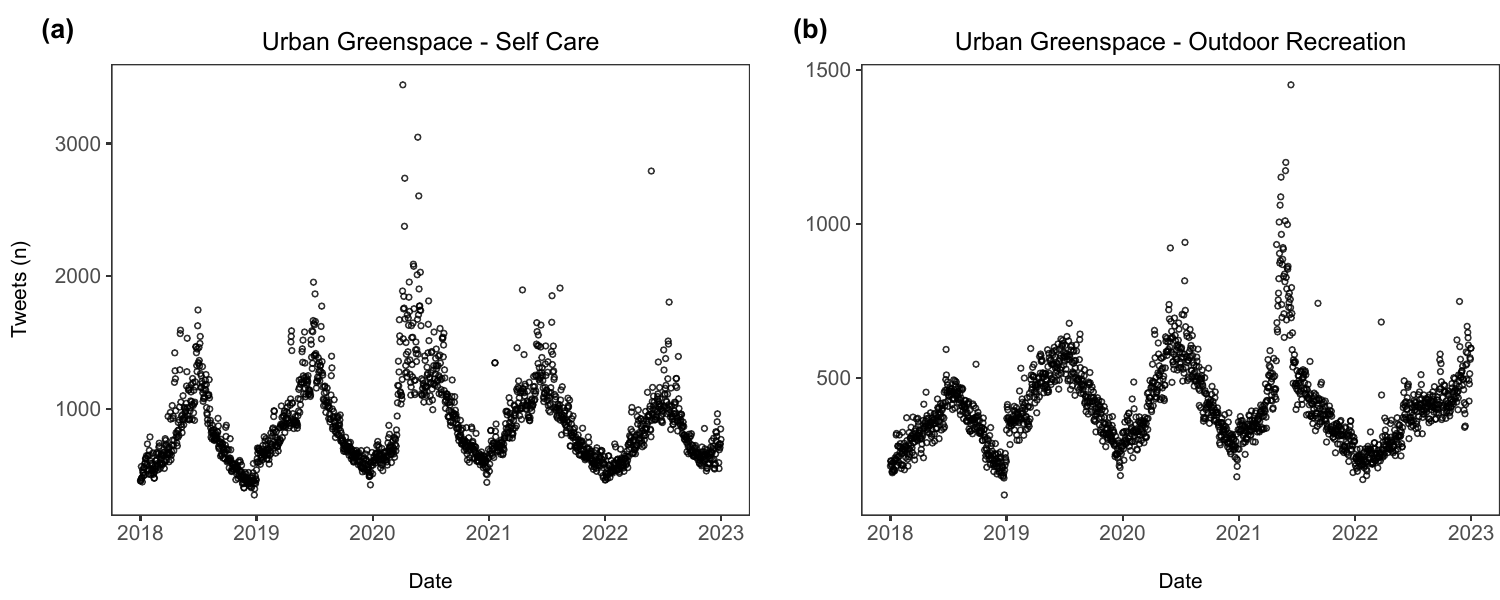}
\caption{Time series of the co-occurrence of \textbf{(a)} urban greenspace features and self care activities and \textbf{(b)} urban greenspace features and outdoor recreation activities in tweets from 2018-2022.}
\label{fig:urban_activities_tweets_ts}
\end{figure}

\section{User plots}\label{sec:user_plots}

\begin{figure}[H]
\centering
\includegraphics[ width=0.85\linewidth]{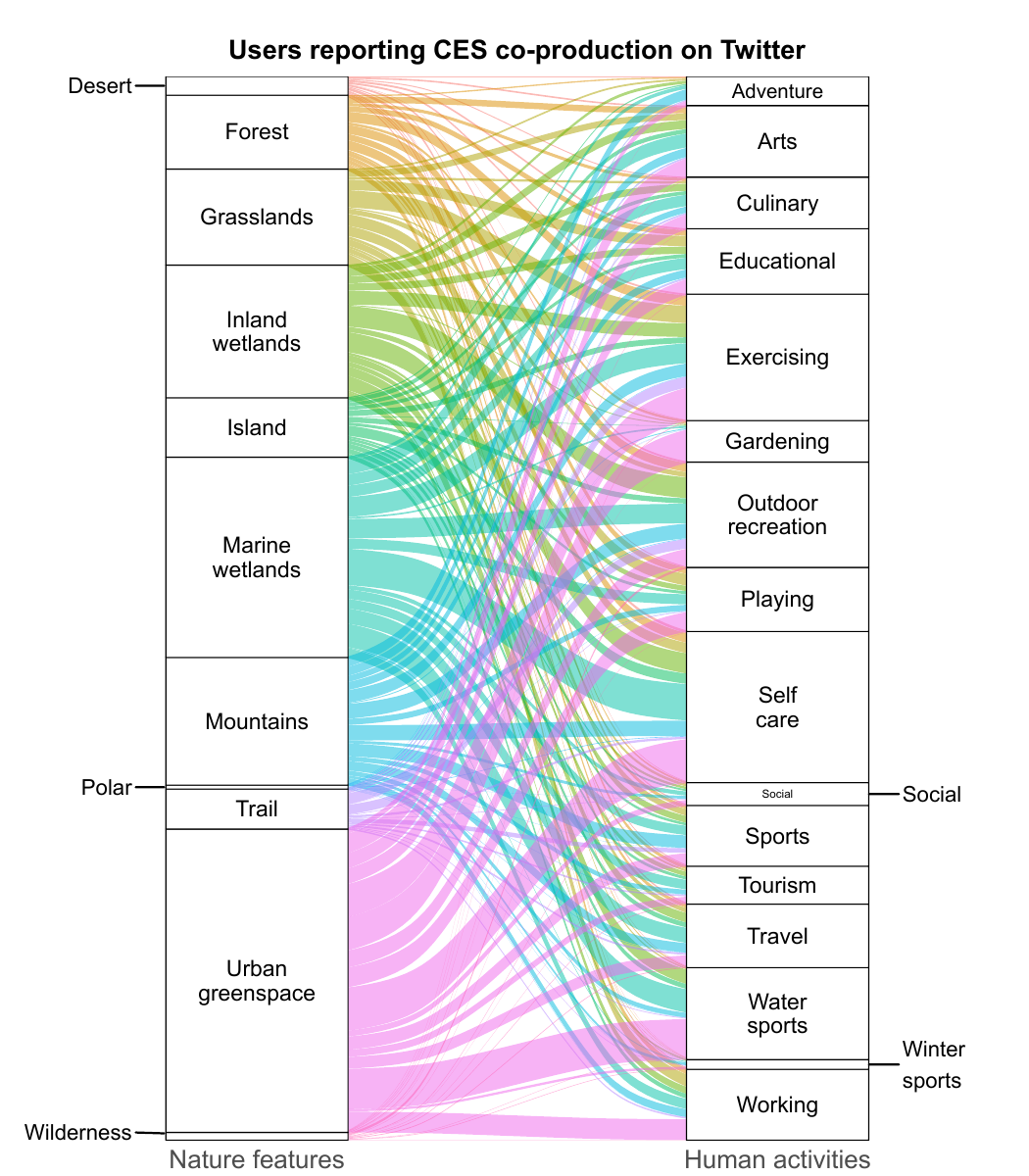}
\caption{Global CES network of users tweeting about acting in nature on Twitter from 2018-2022 (41.7 million tweets). Links are weighted by the number of users tweeting about a given nature feature (grouped by nature class) and human activity (grouped by activity class).}
\label{fig:CES_user_network}
\end{figure}

\begin{figure}[H]
\centering
\includegraphics[ width=0.99\linewidth]{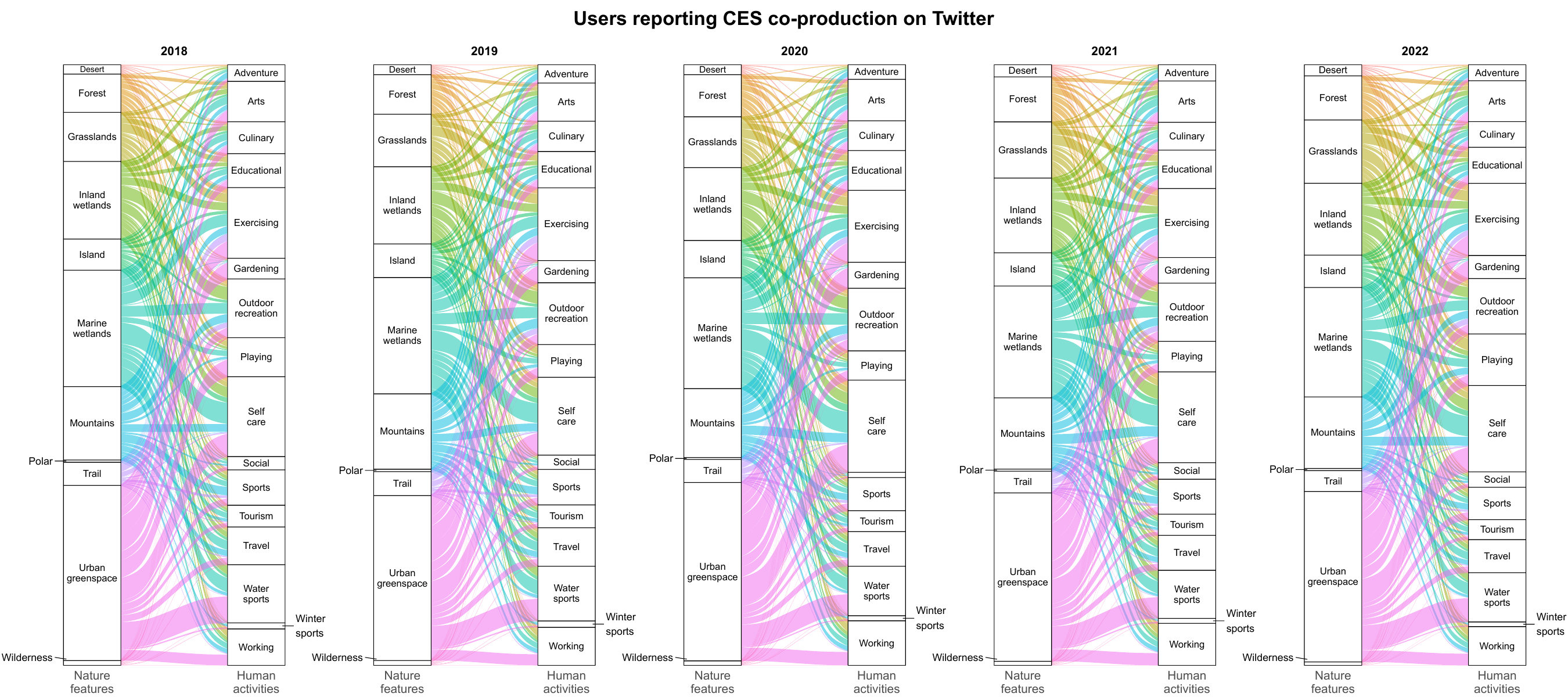}
\caption{Annual CES networks of users tweeting about acting in nature on Twitter from 2018-2022 (41.7 million tweets). Links are weighted by the number of users tweeting about a given nature feature (grouped by nature class) and human activity (grouped by activity class).}
\label{fig:CES_annual_user_network}
\end{figure}

\begin{figure}[H]
\centering
\includegraphics[ width=0.95\linewidth]{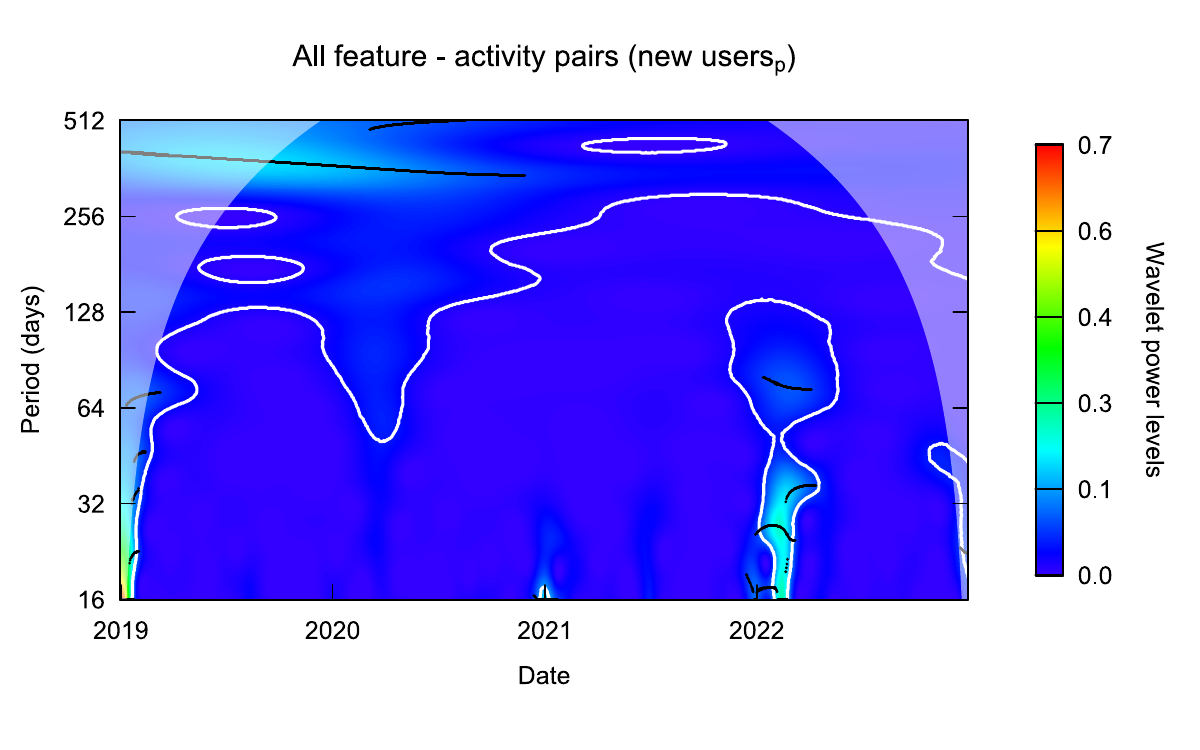}
\caption{Wavelet analysis of users tweeting about the co-occurrence of nature features and human activities from 2019-2022. White contour lines indicate significant time-period domains (95\% significance level). Black lines represent power ridges.}
\label{fig:wavelet_users}
\end{figure}

\begin{figure}[H]
\captionsetup[subfigure]{labelformat=empty}
\begin{subfigure}{0.5\textwidth}
  \centering
  \includegraphics[ width=\linewidth]{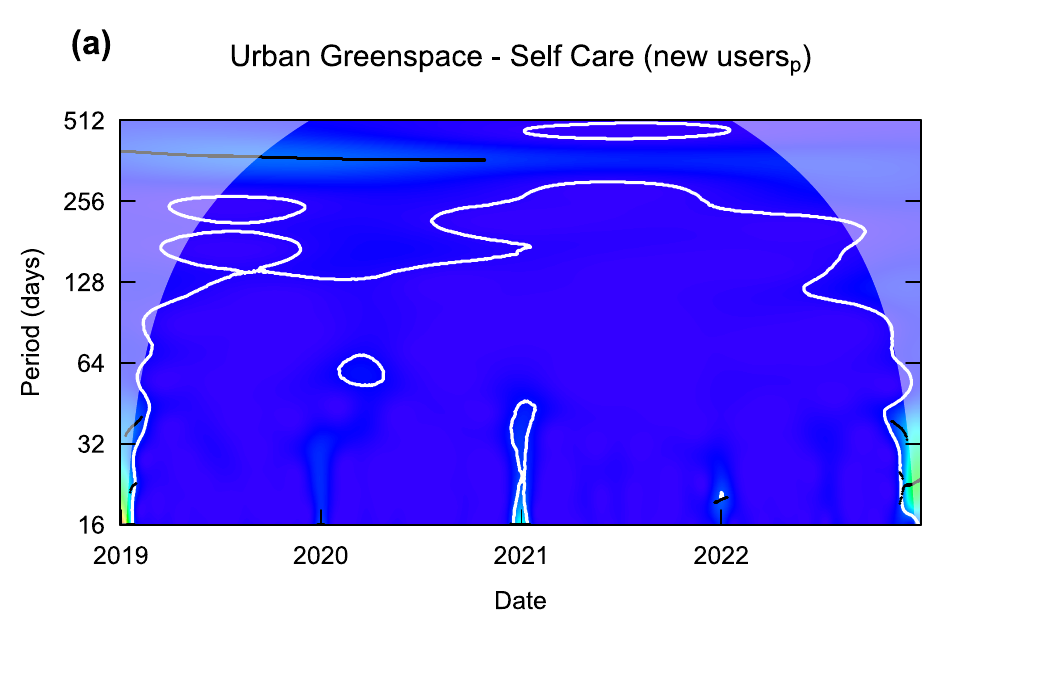}
\end{subfigure}
\begin{subfigure}{0.5\textwidth}
  \centering
  \includegraphics[ width=\linewidth]{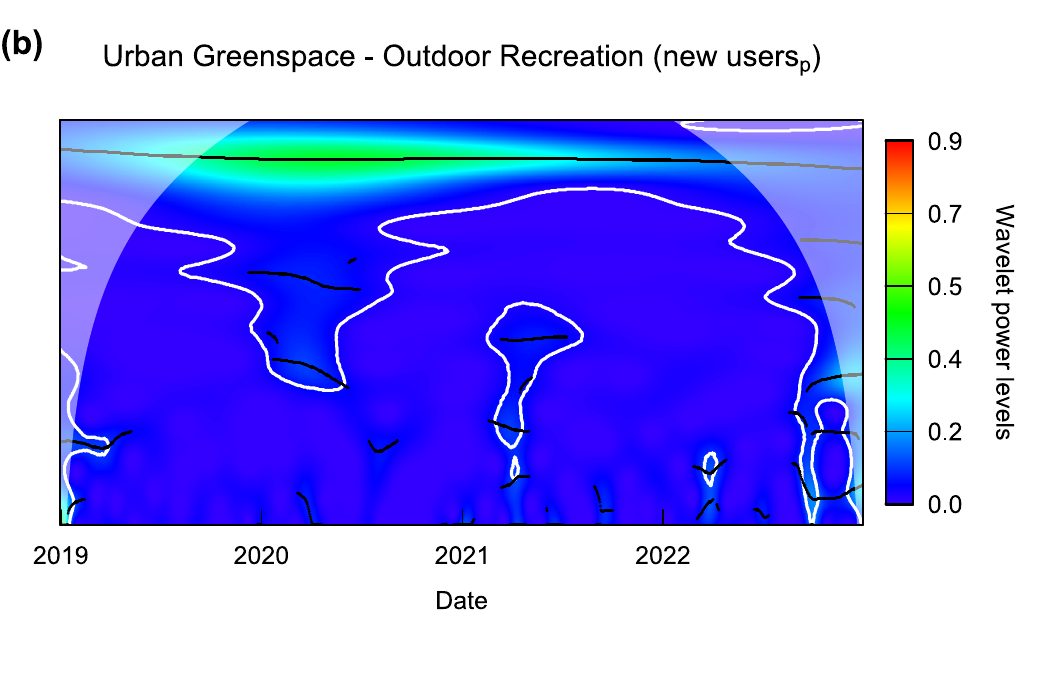}
\end{subfigure}
\caption{Wavelet analysis of the ratio of new users tweeting about the co-occurrence of \textbf{(a)} urban greenspace features and self care activities and \textbf{(b)} urban greenspace features and outdoor recreation activities from 2018-2022. White contour lines indicate significant time-period domains (95\% significance level). Black lines represent power ridges.}
\label{fig:greenspace_user_wavelets}
\end{figure}

\section{Coherency analysis between COVID-19 stringency and CES reported on Twitter} \label{sec:SI_wavelet_analysis}

\begin{figure}[H]
\centering
\includegraphics[ width=0.95\linewidth]{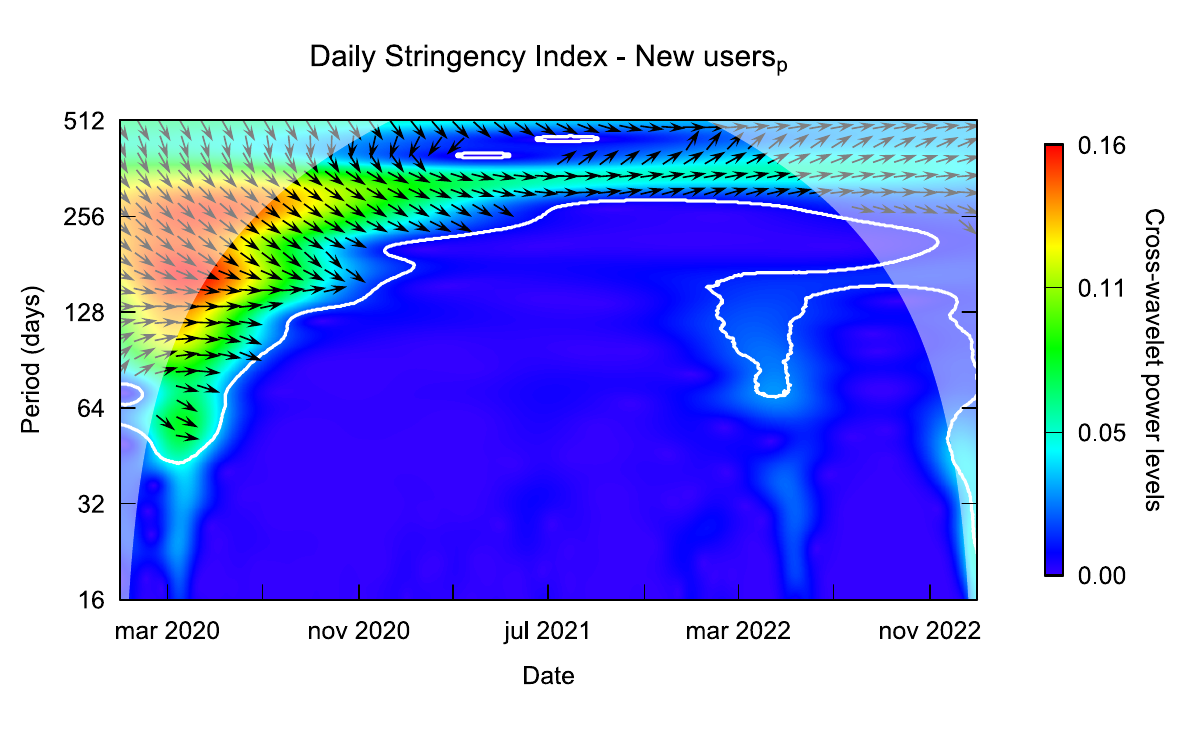}
\caption{Cross wavelet analysis between the COVID-19 stringency index (from \cite{OxCGRT_data}) and the ratio of new users tweeting about CES from 2020-2022. The cross-wavelet spectrogram illustrates the time-frequency relationship between the two time series. The color intensity represents the strength of coherence between the two time series. White contour lines indicate significant time-period domains (95\% significance level). Arrows represent the phase relationship between the series. Right-pointing arrows indicate an in-phase relationship, suggesting synchronized variations between the stringency index and the new user ratio. Left-pointing arrows represent an out-of-phase relationship, suggesting opposite variations.}
\label{fig:SI_users_cross_wavelet}
\end{figure}

\end{appendices}



\begin{thebibliography}{10}
\expandafter\ifx\csname url\endcsname\relax
  \def\url#1{\burl{#1}}\fi
\expandafter\ifx\csname urlprefix\endcsname\relax\def\urlprefix{URL }\fi
\providecommand{\bibinfo}[2]{#2}
\providecommand{\eprint}[2][]{\url{#2}}
\providecommand{\doi}[1]{\url{https://doi.org/#1}}
\bibcommenthead

\bibitem{Kates2011}
\bibinfo{author}{Kates, R.~W.}
\newblock \bibinfo{title}{{What kind of a science is sustainability science?}}
\newblock \emph{\bibinfo{journal}{Proceedings of the National Academy of
  Sciences of the United States of America}} \textbf{\bibinfo{volume}{108}},
  \bibinfo{pages}{19449--19450} (\bibinfo{year}{2011}).

\bibitem{Clark2020}
\bibinfo{author}{Clark, W.~C.} \& \bibinfo{author}{Harley, A.~G.}
\newblock \bibinfo{title}{{Sustainability Science: Toward a Synthesis}}.
\newblock
  \emph{\bibinfo{journal}{https://doi.org/10.1146/annurev-environ-012420-043621}}
  \textbf{\bibinfo{volume}{45}}, \bibinfo{pages}{331--386}
  (\bibinfo{year}{2020}).

\bibitem{Hirons2016}
\bibinfo{author}{Hirons, M.}, \bibinfo{author}{Comberti, C.} \&
  \bibinfo{author}{Dunford, R.}
\newblock \bibinfo{title}{{Valuing Cultural Ecosystem Services}}.
\newblock \emph{\bibinfo{journal}{Annual Review of Environment and Resources}}
  \textbf{\bibinfo{volume}{41}}, \bibinfo{pages}{545--574}
  (\bibinfo{year}{2016}).

\bibitem{TwohigBennett2018}
\bibinfo{author}{Twohig-Bennett, C.} \& \bibinfo{author}{Jones, A.}
\newblock \bibinfo{title}{{The health benefits of the great outdoors: A
  systematic review and meta-analysis of greenspace exposure and health
  outcomes}}.
\newblock \emph{\bibinfo{journal}{Environmental Research}}
  \textbf{\bibinfo{volume}{166}}, \bibinfo{pages}{628--637}
  (\bibinfo{year}{2018}).

\bibitem{Zhang2020}
\bibinfo{author}{Zhang, J.}, \bibinfo{author}{Yu, Z.}, \bibinfo{author}{Zhao,
  B.}, \bibinfo{author}{Sun, R.} \& \bibinfo{author}{Vejre, H.}
\newblock \bibinfo{title}{{Links between green space and public health: A
  bibliometric review of global research trends and future prospects from 1901
  to 2019}} (\bibinfo{year}{2020}).

\bibitem{Fish2016}
\bibinfo{author}{Fish, R.}, \bibinfo{author}{Church, A.} \&
  \bibinfo{author}{Winter, M.}
\newblock \bibinfo{title}{Conceptualising cultural ecosystem services: A novel
  framework for research and critical engagement}.
\newblock \emph{\bibinfo{journal}{Ecosystem Services}}
  \textbf{\bibinfo{volume}{21}}, \bibinfo{pages}{208--217}
  (\bibinfo{year}{2016}).

\bibitem{Raymond2018}
\bibinfo{author}{Raymond, C.~M.}, \bibinfo{author}{Giusti, M.} \&
  \bibinfo{author}{Barthel, S.}
\newblock \bibinfo{title}{{An embodied perspective on the co-production of
  cultural ecosystem services: toward embodied ecosystems}}.
\newblock \emph{\bibinfo{journal}{Journal of Environmental Planning and
  Management}} \textbf{\bibinfo{volume}{61}}, \bibinfo{pages}{778--799}
  (\bibinfo{year}{2018}).

\bibitem{Milcu2013}
\bibinfo{author}{Milcu, A.~I.}, \bibinfo{author}{Hanspach, J.},
  \bibinfo{author}{Abson, D.} \& \bibinfo{author}{Fischer, J.}
\newblock \bibinfo{title}{{Cultural ecosystem services: A literature review and
  prospects for future research}}.
\newblock \emph{\bibinfo{journal}{Ecology and Society}}
  \textbf{\bibinfo{volume}{18}} (\bibinfo{year}{2013}).

\bibitem{Barton2016b}
\bibinfo{author}{Barton, J.}, \bibinfo{author}{Bragg, R.},
  \bibinfo{author}{Pretty, J.}, \bibinfo{author}{Roberts, J.} \&
  \bibinfo{author}{Wood, C.}
\newblock \bibinfo{title}{{The wilderness expedition: An effective life course
  intervention to improve young people’s well-being and connectedness to
  nature}}.
\newblock \emph{\bibinfo{journal}{Journal of Experiential Education}}
  \textbf{\bibinfo{volume}{39}}, \bibinfo{pages}{59--72}
  (\bibinfo{year}{2016}).

\bibitem{Mancini2019}
\bibinfo{author}{Mancini, F.}, \bibinfo{author}{Coghill, G.~M.} \&
  \bibinfo{author}{Lusseau, D.}
\newblock \bibinfo{title}{{Quantifying wildlife watchers’ preferences to
  investigate the overlap between recreational and conservation value of
  natural areas}}.
\newblock \emph{\bibinfo{journal}{Journal of Applied Ecology}}
  \textbf{\bibinfo{volume}{56}}, \bibinfo{pages}{387--397}
  (\bibinfo{year}{2019}).

\bibitem{Lai2019}
\bibinfo{author}{Lai, H.}, \bibinfo{author}{Flies, E.~J.},
  \bibinfo{author}{Weinstein, P.} \& \bibinfo{author}{Woodward, A.}
\newblock \bibinfo{title}{{The impact of green space and biodiversity on
  health}}.
\newblock \emph{\bibinfo{journal}{Frontiers in Ecology and the Environment}}
  \textbf{\bibinfo{volume}{17}}, \bibinfo{pages}{383--390}
  (\bibinfo{year}{2019}).

\bibitem{Palinkas2020}
\bibinfo{author}{Palinkas, L.~A.} \& \bibinfo{author}{Wong, M.}
\newblock \bibinfo{title}{Global climate change and mental health}.
\newblock \emph{\bibinfo{journal}{Current Opinion in Psychology}}
  \textbf{\bibinfo{volume}{32}}, \bibinfo{pages}{12--16}
  (\bibinfo{year}{2020}).

\bibitem{Kossinets2006}
\bibinfo{author}{Kossinets, G.} \& \bibinfo{author}{Watts, D.~J.}
\newblock \bibinfo{title}{Empirical analysis of an evolving social network}.
\newblock \emph{\bibinfo{journal}{Science}} \textbf{\bibinfo{volume}{311}},
  \bibinfo{pages}{88--90} (\bibinfo{year}{2006}).

\bibitem{Gao2016}
\bibinfo{author}{Gao, J.}, \bibinfo{author}{Barzel, B.} \&
  \bibinfo{author}{Barabási, A.-L.}
\newblock \bibinfo{title}{Universal resilience patterns in complex networks}.
\newblock \emph{\bibinfo{journal}{Nature}} \textbf{\bibinfo{volume}{530}},
  \bibinfo{pages}{307–312} (\bibinfo{year}{2016}).

\bibitem{Mateer2021}
\bibinfo{author}{Mateer, T.~J.} \emph{et~al.}
\newblock \bibinfo{title}{Psychosocial factors influencing outdoor recreation
  during the covid-19 pandemic}.
\newblock \emph{\bibinfo{journal}{Frontiers in Sustainable Cities}}
  \textbf{\bibinfo{volume}{3}} (\bibinfo{year}{2021}).

\bibitem{Veldez2020}
\bibinfo{author}{Valdez, D.}, \bibinfo{author}{ten Thij, M.},
  \bibinfo{author}{Bathina, K.}, \bibinfo{author}{Rutter, L.~A.} \&
  \bibinfo{author}{Bollen, J.}
\newblock \bibinfo{title}{Social media insights into us mental health during
  the covid-19 pandemic: Longitudinal analysis of twitter data}.
\newblock \emph{\bibinfo{journal}{J Med Internet Res}}
  \textbf{\bibinfo{volume}{22}}, \bibinfo{pages}{e21418}
  (\bibinfo{year}{2020}).

\bibitem{Hossain2020}
\bibinfo{author}{Hossain, M.~M.}, \bibinfo{author}{Sultana, A.} \&
  \bibinfo{author}{Purohit, N.}
\newblock \bibinfo{title}{Mental health outcomes of quarantine and isolation
  for infection prevention: a systematic umbrella review of the global
  evidence}.
\newblock \emph{\bibinfo{journal}{Epidemiology and Health}}
  \textbf{\bibinfo{volume}{42}} (\bibinfo{year}{2020}).

\bibitem{Venter2021}
\bibinfo{author}{Venter, Z.~S.}, \bibinfo{author}{Barton, D.~N.},
  \bibinfo{author}{Gundersen, V.}, \bibinfo{author}{Figari, H.} \&
  \bibinfo{author}{Nowell, M.~S.}
\newblock \bibinfo{title}{Back to nature: Norwegians sustain increased
  recreational use of urban green space months after the covid-19 outbreak}.
\newblock \emph{\bibinfo{journal}{Landscape and Urban Planning}}
  \textbf{\bibinfo{volume}{214}}, \bibinfo{pages}{104175}
  (\bibinfo{year}{2021}).

\bibitem{Lusseau2023}
\bibinfo{author}{Lusseau, D.} \& \bibinfo{author}{Baillie, R.}
\newblock \bibinfo{title}{Disparities in greenspace access during covid-19
  mobility restrictions}.
\newblock \emph{\bibinfo{journal}{Environmental Research}}
  \bibinfo{pages}{115551} (\bibinfo{year}{2023}).

\bibitem{daMota2020}
\bibinfo{author}{Teles~da Mota, V.} \& \bibinfo{author}{Pickering, C.}
\newblock \bibinfo{title}{{Using social media to assess nature-based tourism:
  Current research and future trends}}.
\newblock \emph{\bibinfo{journal}{Journal of Outdoor Recreation and Tourism}}
  \textbf{\bibinfo{volume}{30}} (\bibinfo{year}{2020}).

\bibitem{Vigl2021}
\bibinfo{author}{Egarter~Vigl, L.} \emph{et~al.}
\newblock \bibinfo{title}{{Harnessing artificial intelligence technology and
  social media data to support Cultural Ecosystem Service assessments}}.
\newblock \emph{\bibinfo{journal}{People and Nature}}
  \textbf{\bibinfo{volume}{3}}, \bibinfo{pages}{673--685}
  (\bibinfo{year}{2021}).

\bibitem{Erskine2021}
\bibinfo{author}{Erskine, E.}, \bibinfo{author}{Baillie, R.} \&
  \bibinfo{author}{Lusseau, D.}
\newblock \bibinfo{title}{{Marine Protected Areas provide more cultural
  ecosystem services than other adjacent coastal areas}}.
\newblock \emph{\bibinfo{journal}{One Earth}} \textbf{\bibinfo{volume}{4}},
  \bibinfo{pages}{1175--1185} (\bibinfo{year}{2021}).

\bibitem{Newman2010}
\bibinfo{author}{Newman, M. E. J. M. E.~J.}
\newblock \emph{\bibinfo{title}{{Networks : An Introduction}}}
  \bibinfo{edition}{1} edn (\bibinfo{publisher}{Oxford University Press},
  \bibinfo{address}{New York}, \bibinfo{year}{2010}).

\bibitem{Kolda2009}
\bibinfo{author}{Kolda, T.~G.} \& \bibinfo{author}{Bader, B.~W.}
\newblock \bibinfo{title}{Tensor decompositions and applications}.
\newblock \emph{\bibinfo{journal}{SIAM Review}} \textbf{\bibinfo{volume}{51}},
  \bibinfo{pages}{455--500} (\bibinfo{year}{2009}).

\bibitem{OxCGRT_data}
\bibinfo{author}{Hale, T.} \emph{et~al.}
\newblock \bibinfo{title}{A global panel database of pandemic policies (oxford
  covid-19 government response tracker)}.
\newblock \emph{\bibinfo{journal}{Nature Human Behaviour 2021 5:4}}
  \textbf{\bibinfo{volume}{5}}, \bibinfo{pages}{529--538}
  (\bibinfo{year}{2021}).

\bibitem{Samuelsson2020}
\bibinfo{author}{Samuelsson, K.}, \bibinfo{author}{Barthel, S.},
  \bibinfo{author}{Colding, J.}, \bibinfo{author}{Macassa, G.} \&
  \bibinfo{author}{Giusti, M.}
\newblock \bibinfo{title}{Urban nature as a source of resilience during social
  distancing amidst the coronavirus pandemic}  (\bibinfo{year}{2020}).

\bibitem{Venter2020}
\bibinfo{author}{Venter, Z.~S.}, \bibinfo{author}{Barton, D.~N.},
  \bibinfo{author}{Gundersen, V.}, \bibinfo{author}{Figari, H.} \&
  \bibinfo{author}{Nowell, M.}
\newblock \bibinfo{title}{Urban nature in a time of crisis: recreational use of
  green space increases during the covid-19 outbreak in oslo, norway}.
\newblock \emph{\bibinfo{journal}{Environmental Research Letters}}
  \textbf{\bibinfo{volume}{15}}, \bibinfo{pages}{104075}
  (\bibinfo{year}{2020}).

\bibitem{SoMe_CES_github}
\bibinfo{author}{Linder, A.~C.} \& \bibinfo{author}{Lusseau, D.}
\newblock \bibinfo{title}{{SoMe-CES}}, \bibinfo{version}{1.0.0}
  (\bibinfo{year}{2023}).
\newblock \bibinfo{note}{\url{https://github.com/annecathrinel/SoMe-CES}}.

\bibitem{Fox2021}
\bibinfo{author}{Fox, N.}, \bibinfo{author}{Graham, L.~J.},
  \bibinfo{author}{Eigenbrod, F.}, \bibinfo{author}{Bullock, J.~M.} \&
  \bibinfo{author}{Parks, K.~E.}
\newblock \bibinfo{title}{{Reddit: A novel data source for cultural ecosystem
  service studies}}.
\newblock \emph{\bibinfo{journal}{Ecosystem Services}}
  \textbf{\bibinfo{volume}{50}} (\bibinfo{year}{2021}).

\bibitem{UDPipe2019}
\bibinfo{author}{Straka, M.} \& \bibinfo{author}{Strakov{\'a}, J.}
\newblock \bibinfo{title}{Udpipe pretrained model - universal dependencies 2.5
  models for {UDPipe} (2019-12-06)} (\bibinfo{year}{2019}).
\newblock \bibinfo{note}{\url{http://ufal.mff.cuni.cz/udpipe}}.

\bibitem{twitter_context_annotations}
\bibinfo{author}{{Twitter Developer Platform}}.
\newblock \bibinfo{title}{twitter-context-annotations} (\bibinfo{year}{2022}).
\newblock
  \bibinfo{note}{\url{https://https://github.com/twitterdev/twitter-context-annotations}}.

\bibitem{activity_catalogue}
\bibinfo{author}{Pospíšil, J.}, \bibinfo{author}{Pospíšilová, H.} \&
  \bibinfo{author}{Trochtová, L.}
\newblock \bibinfo{title}{The catalogue of leisure activities: A new structured
  values and content based instrument for leisure research usable for social
  development and community planning}.
\newblock \emph{\bibinfo{journal}{Sustainability}}
  \textbf{\bibinfo{volume}{14}} (\bibinfo{year}{2022}).

\end{thebibliography}


\end{document}